\documentclass[ALICE,manyauthors]{cernphprep}
\usepackage[comma,square,numbers,sort&compress]{natbib}
\usepackage{hyperref}   
\usepackage{lineno}
\usepackage{xcolor}
\usepackage{xspace}
\usepackage{multirow}
\usepackage[T1]{fontenc} 
\begin{document}
%

\newcommand{\pp}           {pp\xspace}
\newcommand{\ppbar}        {\mbox{$\mathrm {p\overline{p}}$}\xspace}
\newcommand{\XeXe}         {\mbox{Xe--Xe}\xspace}
\newcommand{\PbPb}         {\mbox{Pb--Pb}\xspace}
\newcommand{\pA}           {\mbox{pA}\xspace}
\newcommand{\pPb}          {\mbox{p--Pb}\xspace}
\newcommand{\AuAu}         {\mbox{Au--Au}\xspace}
\newcommand{\dAu}          {\mbox{d--Au}\xspace}
\newcommand{\CuCu}         {\mbox{Cu--Cu}\xspace}

\newcommand{\s}            {\ensuremath{\sqrt{s}}\xspace}
\newcommand{\snn}          {\ensuremath{\sqrt{s_{\mathrm{NN}}}}\xspace}
\newcommand{\pt}           {\ensuremath{p_{\rm T}}\xspace}
\newcommand{\meanpt}       {$\langle p_{\mathrm{T}}\rangle$\xspace}
\newcommand{\ycms}         {\ensuremath{y_{\rm CMS}}\xspace}
\newcommand{\ylab}         {\ensuremath{y_{\rm lab}}\xspace}
\newcommand{\etarange}[1]  {\mbox{$\left | \eta \right |~<~#1$}}
\newcommand{\yrange}[1]    {\mbox{$\left | y \right |~<$~0.5}}
\newcommand{\dndy}         {\ensuremath{\mathrm{d}N_\mathrm{ch}/\mathrm{d}y}\xspace}
\newcommand{\dndeta}       {\ensuremath{\mathrm{d}N_\mathrm{ch}/\mathrm{d}\eta}\xspace}
\newcommand{\avdndeta}     {\ensuremath{\langle\dndeta\rangle}}
\newcommand{\dNdy}           {\ensuremath{\mathrm{d}N_\mathrm{ch}/\mathrm{d}y}\xspace}
\newcommand{\dNdyy}         {\ensuremath{\mathrm{d}N/\mathrm{d}y}\xspace}
\newcommand{\Npart}        {\ensuremath{N_\mathrm{part}}\xspace}
\newcommand{\Ncoll}        {\ensuremath{N_\mathrm{coll}}\xspace}
\newcommand{\dEdx}         {\ensuremath{\textrm{d}E/\textrm{d}x}\xspace}
\newcommand{\RpPb}         {\ensuremath{R_{\rm pPb}}\xspace}
\newcommand{\RAA}         {\ensuremath{R_{\rm AA}}\xspace}

\newcommand{\nineH}        {$\sqrt{s}~=~0.9$~Te\kern-.1emV\xspace}
\newcommand{\seven}        {$\sqrt{s}~=~7$~Te\kern-.1emV\xspace}
\newcommand{\twoH}         {$\sqrt{s}~=~0.2$~Te\kern-.1emV\xspace}
\newcommand{\twosevensix}  {$\sqrt{s}~=~2.76$~Te\kern-.1emV\xspace}
\newcommand{\five}         {$\sqrt{s}~=~5.02$~Te\kern-.1emV\xspace}
\newcommand{\twosevensixnn}{$\sqrt{s_{\mathrm{NN}}}~=~2.76$~Te\kern-.1emV\xspace}
\newcommand{\fivenn}       {$\sqrt{s_{\mathrm{NN}}}~=~5.02$~Te\kern-.1emV\xspace}
\newcommand{\LT}           {L{\'e}vy-Tsallis\xspace}
\newcommand{\GeVc}         {Ge\kern-.1emV$/c$\xspace}
\newcommand{\MeVc}         {Me\kern-.1emV$/c$\xspace}
\newcommand{\TeV}          {Te\kern-.1emV\xspace}
\newcommand{\GeV}          {Ge\kern-.1emV\xspace}
\newcommand{\MeV}          {Me\kern-.1emV\xspace}
\newcommand{\GeVmass}      {Ge\kern-.2emV$/c^2$\xspace}
\newcommand{\MeVmass}      {Me\kern-.2emV$/c^2$\xspace}
\newcommand{\lumi}         {\ensuremath{\mathcal{L}}\xspace}

\newcommand{\ITS}          {\rm{ITS}\xspace}
\newcommand{\TOF}          {\rm{TOF}\xspace}
\newcommand{\ZDC}          {\rm{ZDC}\xspace}
\newcommand{\ZDCs}         {\rm{ZDCs}\xspace}
\newcommand{\ZNA}          {\rm{ZNA}\xspace}
\newcommand{\ZNC}          {\rm{ZNC}\xspace}
\newcommand{\SPD}          {\rm{SPD}\xspace}
\newcommand{\SDD}          {\rm{SDD}\xspace}
\newcommand{\SSD}          {\rm{SSD}\xspace}
\newcommand{\TPC}          {\rm{TPC}\xspace}
\newcommand{\TRD}          {\rm{TRD}\xspace}
\newcommand{\VZERO}        {\rm{V0}\xspace}
\newcommand{\VZEROA}       {\rm{V0A}\xspace}
\newcommand{\VZEROC}       {\rm{V0C}\xspace}
\newcommand{\Vdecay} 	   {\ensuremath{V^{0}}\xspace}

\newcommand{\ee}           {\ensuremath{e^{+}e^{-}}} 
\newcommand{\pip}          {\ensuremath{\pi^{+}}\xspace}
\newcommand{\pim}          {\ensuremath{\pi^{-}}\xspace}
\newcommand{\kap}          {\ensuremath{\rm{K}^{+}}\xspace}
\newcommand{\kam}          {\ensuremath{\rm{K}^{-}}\xspace}
\newcommand{\pbar}         {\ensuremath{\rm\overline{p}}\xspace}
\newcommand{\kzero}        {\ensuremath{{\rm K}^{0}_{\rm{S}}}\xspace}
\newcommand{\lmb}          {\ensuremath{\Lambda}\xspace}
\newcommand{\almb}         {\ensuremath{\overline{\Lambda}}\xspace}
\newcommand{\Om}           {\ensuremath{\Omega^-}\xspace}
\newcommand{\Mo}           {\ensuremath{\overline{\Omega}^+}\xspace}
\newcommand{\X}            {\ensuremath{\Xi^-}\xspace}
\newcommand{\Ix}           {\ensuremath{\overline{\Xi}^+}\xspace}
\newcommand{\Xis}          {\ensuremath{\Xi^{\pm}}\xspace}
\newcommand{\Oms}          {\ensuremath{\Omega^{\pm}}\xspace}
\newcommand{\degree}       {\ensuremath{^{\rm o}}\xspace}
\newcommand{\kstar}        {\ensuremath{\rm {K}^{\rm{* 0}}}\xspace}
\newcommand{\phim}        {\ensuremath{\phi}\xspace}
\newcommand{\pik}          {\ensuremath{\pi\rm{K}}\xspace}
\newcommand{\kk}          {\ensuremath{\rm{K}\rm{K}}\xspace}
\newcommand{\kskm}{$\mathrm{K^{*0}/K^{-}}$}
\newcommand{\phikm}{$\mathrm{\phi/K^{-}}$}
\newcommand{\phixi}{$\mathrm{\phi/\Xi}$}
\newcommand{\phiom}{$\mathrm{\phi/\Omega}$}
\newcommand{\xiphi}{$\mathrm{\Xi/\phi}$}
\newcommand{\omphi}{$\mathrm{\Omega/\phi}$}
\newcommand{\kstf} {K$^{*}(892)^{0}~$}
\newcommand{\phf} {$\mathrm{\phi(1020)}~$}
\newcommand{\dd}{\ensuremath{\mathrm{d}}}
\newcommand{\mT}{\ensuremath{m_{\mathrm{T}}}\xspace}
\newcommand{\krr}{\ensuremath{\kern-0.09em}}
\begin{titlepage}
\PHyear{2021}       
\PHnumber{101}      
\PHdate{28 May}  

\title{Production of K$^{*}(892)^{0}$ and $\phi(1020)$ in \pp and \PbPb collisions at \mbox{\snn $=$ 5.02 \TeV}}
\ShortTitle{K$^{*}(892)^{0}$ and $\phi(1020)$ in \pp and \PbPb
  collisions at \snn $=$ 5.02 \TeV}
\Collaboration{ALICE Collaboration\thanks{See Appendix~\ref{app:collab} for the list of collaboration members}}
\ShortAuthor{ALICE Collaboration} 
\begin{abstract}
The production of K$^{*}(892)^{0}$ and $\phi(1020)$ mesons in
proton--proton (\pp) and lead--lead (\PbPb)  collisions at \snn $=$ 5.02
TeV  has been measured using the ALICE detector at the Large Hadron
Collider (LHC).  The transverse momentum ($p_{\mathrm{T}}$)
distributions of K$^{*}(892)^{0}$ and $\phi(1020)$ mesons  have been
measured at  midrapidity $(|y|<0.5)$ up to $p_{\mathrm{T}}$ $=$ 20
GeV$/c$ in  inelastic pp collisions and for several \PbPb collision
centralities. The collision centrality and collision energy dependence
of the average transverse momenta  agree with  the radial flow
scenario observed with stable hadrons, showing that the effect is
stronger for more  central collisions and higher collision
energies. The $\mathrm{K^{*0}/K}$ ratio is found to be suppressed in
Pb--Pb   collisions relative to pp collisions: this indicates a loss
of the measured K$^{*}(892)^{0}$ signal  due to rescattering of its
decay products in the hadronic phase. In contrast, for the
longer-lived $\phi(1020)$ mesons, no such suppression  is observed. 
The nuclear modification factors~($R_{\rm AA}$) of K$^{*}(892)^{0}$
and $\phi(1020)$ mesons are calculated using pp reference spectra
at the same collision energy. In central Pb--Pb collisions for
\pt~$>$~8~GeV$/c$, the \RAA values of K$^{*}(892)^{0}$ and
$\phi(1020)$ are below unity and observed to be similar to those of
pions, kaons, and (anti)protons. The $R_{\rm AA}$ values at high
$p_{\mathrm T}$ ($>$~8~GeV$/c$) for K$^{*}(892)^{0}$ and $\phi(1020)$ mesons are in
agreement  within uncertainties for $\sqrt{s_\mathrm{NN}}$ $=$ 5.02
and 2.76 TeV.
\end{abstract}
\end{titlepage}

\setcounter{page}{2} 


\section{Introduction}\label{sec:intro} 
Experiments at the Large Hadron Collider (LHC) at CERN have recorded \PbPb
collisions at the center of mass energy \snn $=$ 5.02 TeV, to date the
highest  energy for collisions of heavy ions, that has allowed for the
creation of a long-lived, hot, dense, and strongly interacting QCD
matter~\cite{Shuryak:2014zxa,Andronic:2017pug}. One of the physics interests of ALICE experiment is to study the properties of 
the deconfined  state of quarks and gluons  (the Quark-Gluon Plasma, QGP)  produced in the early stages of the
collision relative to the confined state of hadrons and
resonances (excited state hadrons)~\cite{Gyulassy:2004zy,Shuryak:2008eq,Braun-Munzinger:2015hba}.
In these collisions, several kinds of hadrons and resonances with
different flavors of valence quark content, mass, spin, and lifetime
are produced.  Each of these hadrons and resonances possesses unique 
characteristic features that can be exploited to study the properties
of the medium~\cite{Adams:2005dq}. Strongly decaying resonances like
K$^{*}(892)^{0}$ and $\phi(1020)$ with strange valence quarks have
similar masses and spin $=$ 1,  but different lifetimes of 4.16 $\pm$
0.05 fm$/c$ and 46.3 $\pm$ 0.4 fm$/c$~\cite{Tanabashi:2018oca}, respectively. 
The large difference in the lifetimes of these resonances allows one
to probe the system formed in heavy-ion collisions at different
timescales~\cite{Torrieri:2001ue,Adler:2002sw,Adams:2004ep,Alt:2008iv,Abelev:2008zk,Abelev:2008aa,Aggarwal:2010mt,Adare:2010pt,Anticic:2011zr,Adare:2014eyu,Abelev:2014uua,Adamczyk:2015lvo,Knospe:2015nva,Adam:2017zbf,Acharya:2019qge}.

Experiments usually measure the transverse momentum (\pt), rapidity
($y$), and azimuthal angle  ($\varphi$) distributions of the produced
particles. Other observables are mostly derived from these basic
measurements. The total yields of the resonances like K$^{*}(892)^{0}$
and $\phi(1020)$  dominantly come from the low transverse momentum
(\pt $<$ 3 GeV$/c$) particles  and are sensitive to the rescattering
and regeneration processes in the hadronic phase of the  heavy-ion
collisions~\cite{Adam:2017zbf,Abelev:2014uua,Aggarwal:2010mt,Adams:2004ep,Knospe:2015nva}. Further,
the \pt-integrated yields have been used to construct various particle
ratios to understand strangeness  enhancement in high-energy
collisions~\cite{Adam:2017zbf,Abelev:2014uua,Aggarwal:2010mt,Adams:2004ep,Abelev:2008zk,Abelev:2008aa,Adamczyk:2015lvo,ALICE:2017jyt}.
In the intermediate \pt range (3--6 GeV$/c$), effects of radial flow
and recombination have been probed
\cite{Greco:2003xt,Fries:2003vb}. Different kinds of particle ratios,
particularly  baryon-to-meson, have been used to understand these
dynamics~\cite{Adam:2017zbf,Abelev:2014uua,Adamczyk:2015lvo,Abelev:2007ra,Acharya:2019yoi,Abelev:2014laa,Abelev:2013xaa}.
At high \pt, the phenomenon of energy loss by energetic partons
traversing the dense medium formed in high-energy heavy-ion collisions
has been studied~\cite{Aamodt:2010jd,Acharya:2019yoi,Adam:2015kca,CMS:2012aa,Aad:2015wga,Adams:2003kv,Abelev:2006jr,Adler:2003qi,Rafelski:2001hp,Qin:2015srf}. 
The energy loss process depends on the initial medium density, on the
lifetime of the dense matter, on the path length traversed by the
parton, and on the quark flavor. The contributions of these parameters
can be understood by studying the identified hadron \pt spectra for
various collision centralities and collision energies relative to \pp collisions. 

ALICE has previously measured the K$^{*}(892)^{0}$ and $\phi(1020)$
meson production in \pp and \PbPb collisions at \snn $=$ 2.76
\TeV~\cite{Adam:2017zbf,Abelev:2014uua}.  The low-\pt physics
phenomena of rescattering of resonance decay products and regeneration
of resonances in hadronic medium, radial flow, and strangeness
enhancement are addressed through the measurements of the particle
yield (\dNdyy), yield ratios, and mean transverse momentum (\meanpt). 
The measured \meanpt in central \PbPb collisions is observed to be
15--20\%  higher than in peripheral collisions and is also higher than
the \meanpt measured in nucleus--nucleus collisions
at RHIC energies~\cite{Adams:2004ep,Abelev:2008zk,Abelev:2008aa,Aggarwal:2010mt}, suggesting a stronger radial flow effect at the LHC. In
Ref.~\cite{Acharya:2019yoi}, it is shown that \meanpt  of $\pi$, K,
and p in central Pb--Pb collisions is slightly higher at 5.02 TeV than
at 2.76 TeV. This effect is consistent with the presence of a stronger
radial flow at the highest collision energy in Pb--Pb collisions. The
K$^{*}(892)^{0}$ and $\phi(1020)$ resonances, having a mass similar to
the mass of the proton, can further be used to test this effect.
The \pt-integrated yield of K$^{*}(892)^{0}$ relative to kaons is observed to be suppressed in central \PbPb collisions compared to pp and peripheral \PbPb collisions. No such suppression is observed for the $\phi(1020)$ meson. This suggests that the rescattering of the decay products of the short-lived resonance K$^{*}(892)^{0}$  in the hadronic phase is the mechanism that determines the reduced measurable yield. This characteristic is further supported by the expectations from thermal model predictions for \PbPb collisions with a chemical freezeout temperature of 156 MeV, which does not include rescattering effect~\cite{Stachel:2013zma}.
A detailed study on the energy and system size dependence of \pt-integrated particle yield ratios, $\mathrm{K^{*0}/K}$ and
 $\mathrm{\phi/K}$ is performed. For current measurements, these
 ratios are calculated with the average of particle and anti-particle yields 
 i.e. $( \rm {K}^{\rm{* 0}}~+~\rm  {\overline K}^{\rm{* 0}} ) / \left(
   \rm {K}^{+}~+~\rm {K}^{-} \right)$ and 2$\mathrm{\phi}/ \left( \rm
   {K}^{+}~+~\rm {K}^{-}\right)$, which are denoted as
 $\mathrm{K^{*0}/K}$ and $\mathrm{\phi/K}$, respectively throughout
 this paper unless specified otherwise. The reader is referred to
 Ref.~\cite{Acharya:2019qge} for more elaborate discussions on the
 observation of rescattering in \PbPb collisions, lifetime of the
 hadronic phase and physics related to \pt dependence of particle
 ratios involving K$^{*}(892)^{0}$ and $\phi(1020)$ resonances in
 \PbPb collisions at \snn $=$ 5.02 TeV.

The comparison of K$^{*}(892)^{0}$ and $\phi(1020)$ \pt distributions
to the expected \pt distributions from a blast-wave
function~\cite{Schnedermann:1993ws} with parameters obtained from
combined fits to $\pi^{\mathrm{\pm}}$, K$^{\mathrm{\pm}}$, and
p($\mathrm{\bar{p}}$)~\cite{Abelev:2013vea}, which does not include rescattering 
 effects, shows a suppression of the K$^{*}(892)^{0}$ yield by
 $\approx$40\% for \pt$<$ 3 GeV$/c$. However, it is not yet established
 if the observed \pt-dependence of the K$^{*}(892)^{0}$ suppression is
 only due to the rescattering effect. No such suppression is observed
 for the $\phi(1020)$ meson, suggesting that $\phi(1020)$ mesons
 typically decay outside the fireball (lifetime $\approx$ 10
 fm$/c$~\cite{Aamodt:2011mr}) because of their longer lifetime. 

The high-\pt  parton energy loss is studied by measuring the
nuclear modification factor (\RAA). It is defined as
\begin{equation}
  R_{\mathrm{AA}}=\frac{1}{\langle T_{\mathrm{AA}} \rangle} 
  \frac{\mathrm{d}^{2}N^{\mathrm{AA}}/\left(\mathrm{d}y \mathrm{d}p_{\rm T}\right)}
       {\mathrm{d}^{2}\sigma^{\mathrm{pp}}/\left(\mathrm{d}y \mathrm{d}p_{\rm T}\right)},
       \label{eq33}
\end{equation}

where $\mathrm{d}^{2}N^{\mathrm{AA}}/\left(\mathrm{d}y
  \mathrm{d}p_{\rm T}\right)$ is the yield  of particles in heavy-ion
collisions and $\sigma^{\mathrm{pp}}$ is the production cross section
in pp collisions. $\langle T_{\mathrm{AA}} \rangle$ $=$ $\langle N_{\mathrm{coll}} \rangle/\sigma_{\mathrm{inel}}$ is the average
nuclear overlap function, where $\langle N_{\mathrm{coll}} \rangle$ is the
average number of binary  nucleon--nucleon collisions calculated using
MC Glauber~\cite{Miller:2007ri} simulations, and
$\sigma_{\mathrm{inel}}$ is the inelastic pp cross
section equal to $\mathrm{(67.6\pm 0.6)}$ mb at \snn~$=$~5.02~TeV~\cite{Loizides:2017ack,Acharya:2019yoi}.
The \RAA
measurements in \PbPb collisions at
\snn~$=$~2.76~TeV~\cite{Adams:2004ep,Adam:2015kca} show that at high
transverse momentum (\pt $>$ 8 GeV$/c$) nuclear modification factors
for $\pi$, K, p, K$^{*}(892)^{0}$, and $\phi(1020)$ are consistent within uncertainties. 
This suggests that the partonic energy loss in the dense medium produced in heavy-ion collisions does not change the relative particle abundancies at $p_{\mathrm T}$ $>$~8~GeV$/c$ in the light quark sector (u, d, s).
The new measurements in \pp and \PbPb collisions at
\snn~$=$~5.02~TeV will be useful in the study of energy dependence of \RAA
and in further testing the flavor dependence of partonic energy loss
in the dense medium produced in heavy-ion collisions. 
The centrality, collision energy and flavor dependence of the \RAA are
studied using precise measurements at the highest beam energy
available for Pb--Pb collisions at the LHC.

In this article, K$^{*}(892)^{0}$ and $\phi(1020)$ meson production is 
studied at midrapidity, $|y|$ $<$ 0.5, over a wide transverse momentum
range up to 20~GeV$/c$ in \PbPb and inelastic \pp collisions at \snn $=$ 5.02 TeV. 
Throughout this article, the results for K$^{*}(892)^{0}$ and  
$\mathrm{\overline{K}}^{*}(892)^{0}$ are averaged and denoted by the 
symbol \kstar, and $\phi(1020)$ is denoted by $\phim$ unless specified 
otherwise. The article is organized as follows: Section~\ref{sec:analysis} 
describes the data analysis techniques, which include the event and 
track selection, the technique adopted to obtain the yields of the resonances,
correction factors and systematic uncertainties. Section~\ref{sec:results} 
presents results related to the K$^{*0}$ and $\phi$ meson $p_{\mathrm{T}}$ 
spectra, yields, mean transverse momentum, particle ratios and nuclear
modification factors. A summary of the work presented in the article
is given in Section~\ref{sec:conc}.

\section{Data analysis}\label{sec:analysis}
 The measurements of \kstar and \phim meson production in \pp and
 \PbPb collisions at \snn $=$ 5.02 \TeV have been performed on data
 taken with the ALICE detector in the year 2015. The resonances are
 reconstructed via their hadronic decay channels, \kstar$\rightarrow
 \pi^{\pm}$K$^{\mp}$ (B.R. $=$ 66.6\%~\cite{Tanabashi:2018oca})  and
 \phim$\rightarrow$ K$^{+}$K$^{-}$ (B.R. $=$ 49.2\%~\cite{Tanabashi:2018oca}). 
In pp collisions, \kstar and \phim mesons are measured in inclusive
inelastic events, whereas in \PbPb collisions they are measured in
eight collision centrality classes 0--10\%, 10--20\%, 20--30\%,
30--40\%, 40--50\%, 50--60\%, 60--70\%, and 70--80\%~\cite{Abelev:2013qoq}. 

\subsection{Event and track selection}
A detailed description of the ALICE detector can be found in
Refs.~\cite{Aamodt:2008zz, Abelev:2014ffa}. The measurements are
obtained with the ALICE central barrel detectors, which are located
inside  a solenoidal magnet providing a magnetic field of 0.5 T, and
are used for tracking, particle identification and reconstruction of
the primary vertex. The measurements have been performed by using
central barrel detectors: the Inner Tracking System (\ITS), the Time
Projection Chamber (\TPC), and the Time-of-Flight (\TOF)
detector. These detectors have full azimuthal coverage around
midrapidity, at pseudorapidity $|\eta|~\textless$~0.9.  
The primary vertex position is determined from global tracks~\cite{Abelev:2014ffa}. 
Global tracks are reconstructed using both the TPC and ITS, and are
used to determine the primary vertex position~\cite{Abelev:2014ffa}.  
Events are selected according to the position of the primary vertex along the beam axis 
($\mathrm{v}_{\mathrm{z}}$), which is required to be within 10 cm 
from the nominal interaction point to ensure a uniform acceptance and
reconstruction efficiency in the pseudorapidity region
$|\eta|~\textless$~0.8. In addition, the difference between the
vertices reconstructed with the two innermost layers of the ITS and
those reconstructed with global tracks ($| \mathrm{v}_{\mathrm{z
    Track}} - \mathrm{v}_{\mathrm{z SPD}}|$) is required to be less than 0.5 cm.
This selection is required to reject pile-up events in pp collisions,
which are less than 1\% of the overall number of events.
\PbPb collisions have negligible pile-up.  A pair of scintillator arrays (\VZERO detector) that cover the pseudorapidity region 2.8~$\textless~\eta~\textless$~5.1 
 (\VZERO-A) and -3.7~$\textless~\eta~\textless$~-1.7 (\VZERO-C), 
is used for the interaction trigger both in \pp and in \PbPb
collisions. The trigger is defined as a coincidence between the  
\VZERO-A and the \VZERO-C. In addition,  at least one hit in the central barrel
detector SPD is required for the minimum bias trigger in \pp collisions.
The \VZERO detector signal is the total charge collected (V0M amplitude) in the
detector, which is proportional to the charged particle multiplicity in
its acceptance, and it is used to classify the \PbPb events into
centrality classes, defined in terms of percentiles of the hadronic
cross section.  A Glauber Monte Carlo model is fitted to the \VZERO
amplitude distribution to compute the fraction of the hadronic cross
section corresponding to any given range of \VZERO amplitudes. Based
on these studies, the data are divided into several centrality classes
\cite{Adam:2015ptt}. The number of events analyzed after the event
selections are $\approx$ 110$\times 10^{6}$ and $\approx$ 24$\times
10^{6}$ in minimum bias pp and Pb--Pb collisions, respectively.

 \kstar and \phim mesons are reconstructed using global tracks.
To ensure high tracking efficiency and to limit the contamination due to
secondary particles and tracks with wrongly associated hits, global
selected tracks are required to have a minimum number of TPC hits
associated to the track (70 out of a maximum of 159). The reconstructed track $\chi^{2}$ normalized to the number of TPC clusters is required to be lower than 4.
To reduce the contamination from beam-background events and secondary
particles coming from weak decays, selection criteria on the distance
of closest approach to the primary vertex in the transverse plane
($DCA_{xy}$) of the selected tracks and in the beam direction ($DCA_{z}$) are
applied~\cite{Abelev:2014ffa}. The value of $DCA_{xy}$ is required to
be $DCA_{xy}$ (\pt ) $\textless~0.0105 + 0.035\pt^{-1.1}$ cm (\pt in GeV$/c$) which
corresponds to 7 times the $DCA_{xy}$ resolution, and $DCA_{z}$ is
required to be less than 2 cm. The \pt is requested to be larger than 0.15~GeV$/c$.
The charged tracks are selected within the pseudorapidity range
$|\eta|~\textless$~0.8, which ensures uniform acceptance and the best
reconstruction efficiency. Furthermore, the charged tracks from the
decay of weakly decaying kaons are rejected.

The \TPC and \TOF are used to identify pions and kaons 
by measuring the specific ionization energy loss (\dEdx) 
in the \TPC and their time-of-flight in the \TOF, respectively. 
Whenever the TOF information for a given track is not available, only
the TPC information is used for particle identification.
The \dEdx resolution of the \TPC is denoted as $\sigma_{\mathrm{\TPC}}$. 
For \kstar and \phim meson reconstruction in \PbPb collisions, and \kstar reconstruction in pp collisions,
pion and kaon candidates 
are required to have  $\langle\dEdx\rangle$ within 
 2$\sigma_{\mathrm{\TPC}}$ of the expected \dEdx values for each
 particle species over the whole momentum range. To further reduce the
 number of misidentified particles, the measured time-of-flight is
 required not to deviate from the expected value for each given mass
 hypothesis by more than 3$\sigma_{\mathrm{\TOF}}$
 ($\sigma_{\mathrm{\TOF}}$ $\approx$ 60 ps)~\cite{Acharya:2019yoi}. For
 \phim meson reconstruction in pp collisions, the kaon candidates are selected using the
 TPC with selection criteria of  6$\sigma_{\mathrm{\TPC}}$,
 4$\sigma_{\mathrm{\TPC}}$ and 2$\sigma_{\mathrm{\TPC}}$ on the
 measured $\langle\dEdx\rangle$ distributions in the momentum ranges $p<$ 0.3 \GeVc, 0.3 $<p<$ 0.4 \GeVc, and $p>$ 0.4 \GeVc,
 respectively. In addition, a 3$\sigma_{\mathrm{\TOF}}$ selection
 criterion is applied on the time-of-flight over the measured momentum
 range whenever the \TOF information is available.

\subsection{Yield extraction}
\begin{figure}[tb]
    \begin{center}
    \includegraphics[width = 0.7\textwidth]{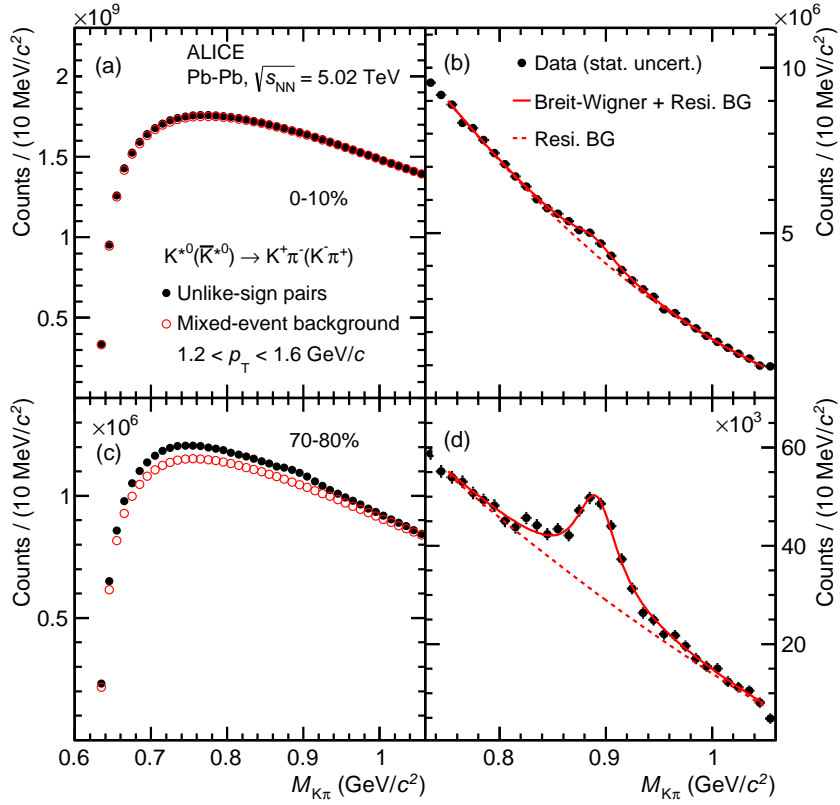}
    \end{center}
    \caption{Invariant-mass distributions of \pik pairs for the 0--10\% and
      70--80\% centrality classes in \PbPb collisions at \snn~$=$~5.02 \TeV for the transverse
      momentum range 1.2 \textless~\pt~\textless~1.6 \GeVc. The left
      panels show the unlike charge \pik invariant-mass distribution
      from the same event and the normalized mixed event background. The right panels report the invariant-mass distribution after subtraction of the combinatorial background for \kstar. The solid curves represent fits to the distributions and the dashed curves are the components of the fits that describe the residual background. The statistical uncertainties are shown by bars.}
    \label{fig1:KstarInv}
\end{figure}

\begin{figure}[tb]
    \begin{center}
   \includegraphics[width = 0.7\textwidth]{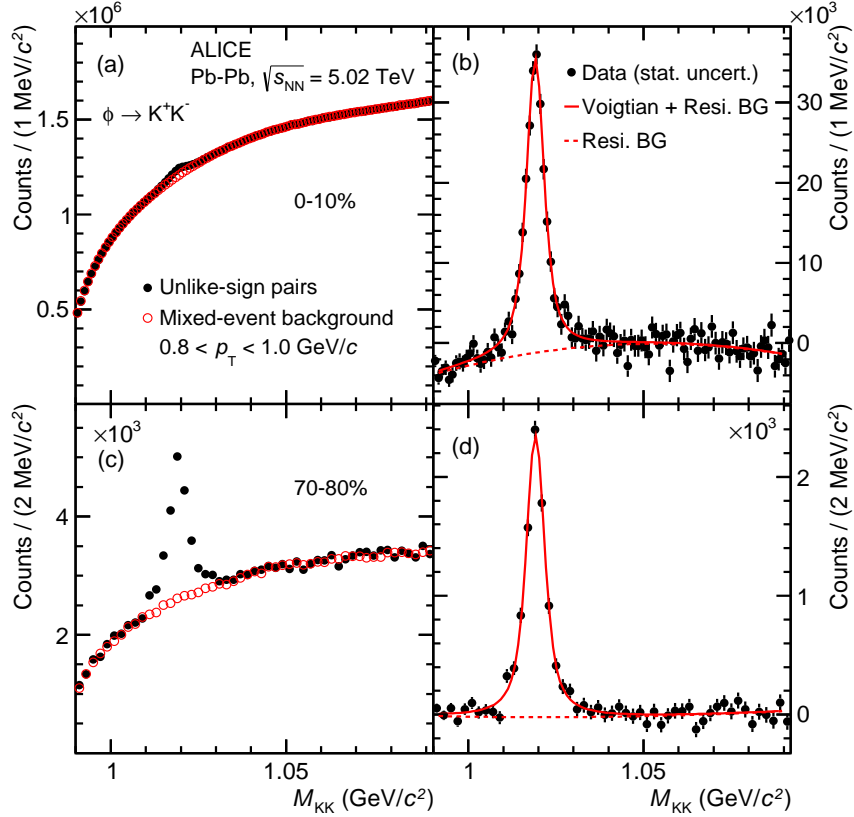}
    \end{center}
    \caption{Invariant-mass distributions of \kk pairs for the 0--10\% and
      70--80\% centrality classes in \PbPb collisions at \snn~$=$~5.02
      \TeV for the transverse momentum ranges 0.8~\textless~\pt~\textless~1.0 \GeVc. The left panels show the unlike charge \kk
      invariant-mass distribution from the same event and the normalized mixed event background. The right panels report the invariant-mass distribution after subtraction of the combinatorial background for \phim. The solid curves represent fits to the distributions and the dashed curves are the components of the fits that describe the residual background. The statistical uncertainties are shown by bars.}
    \label{fig:PhiInv}
  \end{figure}
  
The raw yields are extracted in each \pt bin and centrality class as done in previous 
  work~\cite{Abelev:2012hy,Acharya:2018orn,Adam:2016bpr,Abelev:2014uua,Adam:2017zbf}. The
  \pt spectra for \kstar (\phim) mesons cover the  range 0--20 GeV$/c$
  (0.4--20 GeV$/c$) for pp collisions. For \PbPb collisions, the \pt
  spectra of \kstar and \phim mesons are measured from \pt $=$
0.4 GeV$/c$ up to 20 GeV$/c$ in all centrality classes. 
The \kstar and \phim mesons are reconstructed via their hadronic decay
channels by calculating the invariant mass of their decay
daughters. For each event, the unlike-sign kaons and pions are paired
for \kstar, and unlike-sign kaons are paired for the \phim meson to
construct the invariant-mass distribution. The rapidity of the
daughters pair is required to lie in the range, $|y|$ $<$ 0.5. An
event mixing technique is used to estimate the combinatorial
background where the kaons and pions from one event are mixed with
oppositely charged kaons and pions from other events. 
 Two events are mixed only if they have similar multiplicity ($|\Delta
n|~\textless~\rm{5}$) and collision vertex ($|\Delta
\rm{v_{z}}|~\textless~\mathrm{1~cm}$). To reduce the
statistical uncertainties from the background distribution, each event
is mixed with five other events. Then the mixed-event
invariant mass distribution is normalized in the mass region outside
of the mass peak, namely 1.1 $<\mathrm{M}_{\mathrm{K}\pi}<$ 1.15
GeV$/c^{2}$ and 1.035 $<\mathrm{M_{KK}}<$ 1.045 GeV$/c^{2}$ for \kstar
and \phim mesons, respectively. The left panels of Figs.~\ref{fig1:KstarInv}
and~\ref{fig:PhiInv} show the  invariant mass distributions of
unlike-sign K$\pi$ and KK pairs from the same event (black marker) and
the normalized mixed event background (red marker) for the transverse
momentum ranges 1.2~\textless~\pt~\textless~1.6~\GeVc and  0.8~\textless~\pt~\textless~1.0~\GeVc, respectively.  The invariant mass distributions are shown for
the 0--10\% and 70--80\% centrality classes in Pb--Pb collisions.  The
combinatorial background subtracted invariant-mass distributions are
fitted using a combined function to describe the signal peak and the
residual background. As shown in the right panels of
Figs.~\ref{fig1:KstarInv} and ~\ref{fig:PhiInv}, respectively, a
Breit-Wigner function (Eq.~\ref{kstarfitfun}) is used to describe the
\kstar peak and a Voigtian function  (a convolution of a Breit-Wigner
and a Gaussian function, Eq.~\ref{phifitfun})  is used to describe
the \phim peak.  A second order polynomial is used to describe the
residual background in both cases. The residual background is what remains after
the combinatorial background subtraction and it is mainly due to
correlated pairs from real resonance decays where the daughter
particles are misidentified as K or $\pi$. 
 The signal peak fit functions for \kstar and \phim are
\begin{equation}
      \frac{\mathrm{d}N}{\mathrm{d}M_{\mathrm{K}\pi}} = \frac{N_{\mathrm{raw}}}{2\pi}
       \frac{\Gamma} {(M_{\mathrm{K}\pi} - M_0)^{2} + \frac{\Gamma^{2}}{4}},
      \label{kstarfitfun}
    \end{equation}

\begin{equation}
\frac{\mathrm{d}N}{\mathrm{d}M_\mathrm{KK}} =
\frac{N_{\mathrm{raw}}}{2\pi}\int \frac{\Gamma} {(M_
  \mathrm{KK}-m^{'})^2+\Gamma^2/4} \frac{e^{{-(m^{'} -M_{0})}^{2}/2\sigma^2}}{\sqrt{2\pi}\sigma} dm^{'},
\label{phifitfun}
\end{equation}

where $M_{\mathrm{K}\pi}$ and $M_{\mathrm{KK}}$ are the reconstructed invariant masses 
of $\mathrm{K}^{*0}$ and $\phi$ mesons. $M_{\mathrm{0}}$, $\Gamma$,
and $N_{\mathrm{raw}}$  are the mass, width and raw yield of the
resonances, respectively. The parameter $\sigma$ in
Eq.~\ref{phifitfun} represents the mass resolution, which depends on
\pt. The widths of \kstar and \phim are fixed to the vacuum
values~\cite{Tanabashi:2018oca} while fitting the invariant mass
distributions. For the \phim meson, the $\sigma$ is kept free. The measured $\sigma$ on the \phim mass is \pt dependent,  varies between 1--2~MeV$/c^{2}$ and its
 values are consistent with the values obtained from Monte Carlo simulations.
 The raw particle yields are extracted by integrating the invariant
 mass distribution within the mass interval approximately
 $M_{0}\pm2\Gamma$ and subtracting the integral of the residual
 background function in the same mass region. The resonance yields
 beyond the integration region are obtained by integrating the tail
 part of the signal fit function; these yields are then added to the
 yields extracted by integrating the invariant mass distribution.

\subsection{Yield correction}
 
\subsubsection{Acceptance and reconstruction efficiency}
The raw transverse momentum distributions are corrected for the detector acceptance 
and reconstruction efficiency ($A \times \epsilon_{rec}$). These correction factors are 
evaluated by using Monte Carlo (MC) events generated with PYTHIA8 
(Monash 2013 Tune)~\cite{Skands:2014pea} for \pp and HIJING~\cite{Wang:1991hta} 
for \PbPb collisions,  and by transporting the particles through a
full simulation of the ALICE detector with {\sc GEANT3}~\cite{Brun:1082634}. 
The $A \times \epsilon_{rec}$  has a centrality dependence in \PbPb collisions and a deviation of $\approx$ 5--7\% is observed from the most central to the
most peripheral centrality classes. As the real data and the generated
MC spectral shapes are different, the \pt spectra of the generated mesons are re-weighted to the respective \pt spectra from the data in an iterative method 
to re-estimate the efficiency~\cite{Acharya:2019bli}.  
The effect of re-weighting the generated \pt spectra on $A \times
\epsilon_{rec}$  is $\approx$ 4--6\% at low-\pt ($<1$~\GeVc) and is
negligible at high \pt ($>1$~\GeVc). 

\subsubsection{Normalization}
The normalized yield is given by
\begin{equation}
\frac{1}{N_{\mathrm{event}}}~\frac{\mathrm{d^{2}}N}
{\mathrm{d}y\mathrm{d}\pt} =
\frac{1}{N_{\mathrm{event}}^{\mathrm{acc}}}
\frac{N_{\mathrm{raw}}}{\Delta y\Delta\pt}
\frac{\epsilon_{\mathrm{trig}} ~ \epsilon_{\mathrm{vert}} ~
  \epsilon_{\mathrm{SL}}}{(A \times \epsilon_{rec}) ~ BR},
\label{nor}
\end{equation}

where $\Delta y$ and $\Delta\pt$ are the widths of rapidity and \pt
bins, respectively. The raw spectra are corrected for the branching
ratio ($BR$). The extracted yields are normalized to the number of
analyzed events ($N_{\mathrm{event}}^{\mathrm{acc}}$). In order to
obtain the absolute resonance yields per inelastic pp collision, the
factor $\epsilon_{\mathrm{trig}}$ $=$ 0.757 $\pm$ 0.019 due to trigger
efficiency is used. This is the ratio between the V0 visible cross
section~\cite{ALICE-PUBLIC-2016-005} and the inelastic cross
section~\cite{Loizides:2017ack}.  

The correction factor $\epsilon_{\mathrm{vert}}$ accounts for the
vertex reconstruction efficiency, which is calculated as the ratio of
the number of events having good vertex to the total number of
triggered events. This is estimated to be 0.958 in \pp collisions.
The signal loss correction, $\epsilon_{\mathrm{SL}}$ accounts for the
loss in \kstar and \phim yields that is caused by the event selection
with minimum bias trigger, rather than all inelastic events. The
$\epsilon_{\mathrm{SL}}$ has a \pt dependence and is only significant
for low \pt ($<$ 2.5 \GeVc). It is calculated as the ratio of the \pt
spectrum from inelastic events to the \pt spectrum from triggered events. The
value of  $\epsilon_{\mathrm{SL}}$ is less than 1.05 for both \kstar
and \phim mesons in pp collisions. The effects of inelastic trigger,
vertex reconstruction efficiency and signal loss corrections are
negligible in \PbPb collisions~\cite{Abelev:2013vea, Abelev:2014laa}
and, hence, are not considered.  

\subsection{Systematic uncertainties}
\label{sec_syst}
The systematic uncertainties in the measurement of \kstar and \phim production in \pp 
and \PbPb collisions at \snn $=$ 5.02 \TeV have been estimated by considering 
uncertainties due to signal extraction, track selection criteria,
particle identification, global tracking efficiency, uncertainty in
the material budget of the ALICE apparatus and the hadronic
interaction cross section in the detector material. To study the
systematic uncertainty for \kstar and \phim in \PbPb and \kstar in pp,
an approach similar to that described
in Refs.~\cite{Abelev:2012hy,Adam:2016bpr,Abelev:2014uua,Adam:2017zbf} has
been adopted. For the estimation of systematic uncertainties for \phim
in pp, a similar approach is followed as in
Refs.~\cite{Acharya:2019wyb,Acharya:2020uxl}.  

A summary of the systematic uncertainties from various sources, for 
\kstar and \phim in \pp and \PbPb collisions is given
in Ref.~\cite{Acharya:2019qge} where the values of relative systematic
uncertainties are quoted for  low,  intermediate, and high \pt.
The track selection criteria have been varied to study the systematic effect
due to the track selection. In order to study the effect of the choice
of particle identification criteria of the daughter tracks on raw
yield extraction, the selection criteria on TPC and TOF have been
varied. To estimate the systematic uncertainty of particle identification
(PID), the $N\sigma_{\rm{TPC/TOF}}$ cut is varied by
1$\sigma_{\rm{TPC/TOF}}$ from the default PID selection criterion.

The uncertainty due to the signal extraction includes variations of
the event mixing background normalization range, signal fit range,
residual background fit function, choice of combinatorial background,
and mass resolution. The mixed event background distributions for
\kstar and \phim have been normalized in different invariant-mass
regions excluding the signal peaks; the change in yield is
considered as the systematic uncertainty. 
The K$\pi$ invariant-mass fitting ranges are varied by 10--50
MeV$/c^{2}$ for \kstar whereas for the \phim the KK invariant-mass
fitting ranges are varied by 5--10 MeV$/c^{2}$.
The residual background is fitted with a third-order polynomial for
\PbPb collisions, and in pp collisions, a first- and third-order
polynomial is used for systematic studies. The systematic
uncertainties due to the combinatorial background are estimated by
changing the method of background reconstruction (like sign and event
mixing). 

Another source of uncertainty comes from the determination of the
global tracking efficiency, which arises from the  ITS-TPC track
matching efficiency.  The systematic uncertainties due to global
tracking efficiency are calculated from the corresponding values for
single charged particles uncertainty and by combining the two charged
tracks used in the invariant mass reconstruction of \kstar and \phim.
In both \pp and \PbPb collisions, this contribution has been estimated
to be \pt dependent for charged particles~\cite{Abelev:2014laa}.  

The material budget of the ALICE detector setup is known with an uncertainty of 7\% in terms of radiation length, determined
on the basis of $\gamma$ conversion measurements~\cite{Aamodt:2010dx}. 
The systematic uncertainty
contribution due to material budget is thus estimated by varying the
amount of material by $\pm$ 7\% in the Monte Carlo simulation. 
The systematic uncertainty due to the hadronic interaction cross
section in the detector material is estimated by comparing different
transport codes: {\sc GEANT3}~\cite{Brun:1082634}, {\sc GEANT4}~\cite{Agostinelli:2002hh}, and {\sc FLUKA}~\cite{Battistoni:2007zzb}.
The effects of material budget and hadronic interactions are evaluated
by combining the uncertainties for a pion and a kaon (in case of
\kstar), and for two kaons (in case of \phim) according to the
kinematics of the decay~\cite{Abelev:2014laa}.  These effects
are found to be negligible at intermediate and high \pt for both
\kstar and \phim. 

Raw yield extraction and global tracking efficiency dominate total
uncertainties in the lowest and highest \pt intervals. The total
systematic uncertainties for \kstar and \phim amount to 10.9--12.3\%
(9.1--13.0\%)  and 6.4--9.2\% (5.4--9.5\%) in Pb--Pb (pp) collisions,
respectively. Among the sources of systematic uncertainty, the yield
extraction is the only fully uncorrelated source, while track
selection, PID, global tracking efficiency, material budget and
hadronic interaction are correlated across different centrality
classes.

\section{Results}\label{sec:results} 
\subsection{\pt spectra in \pp collisions}
\begin{figure}[tb]
    \begin{center}
 \includegraphics[scale = 0.7]{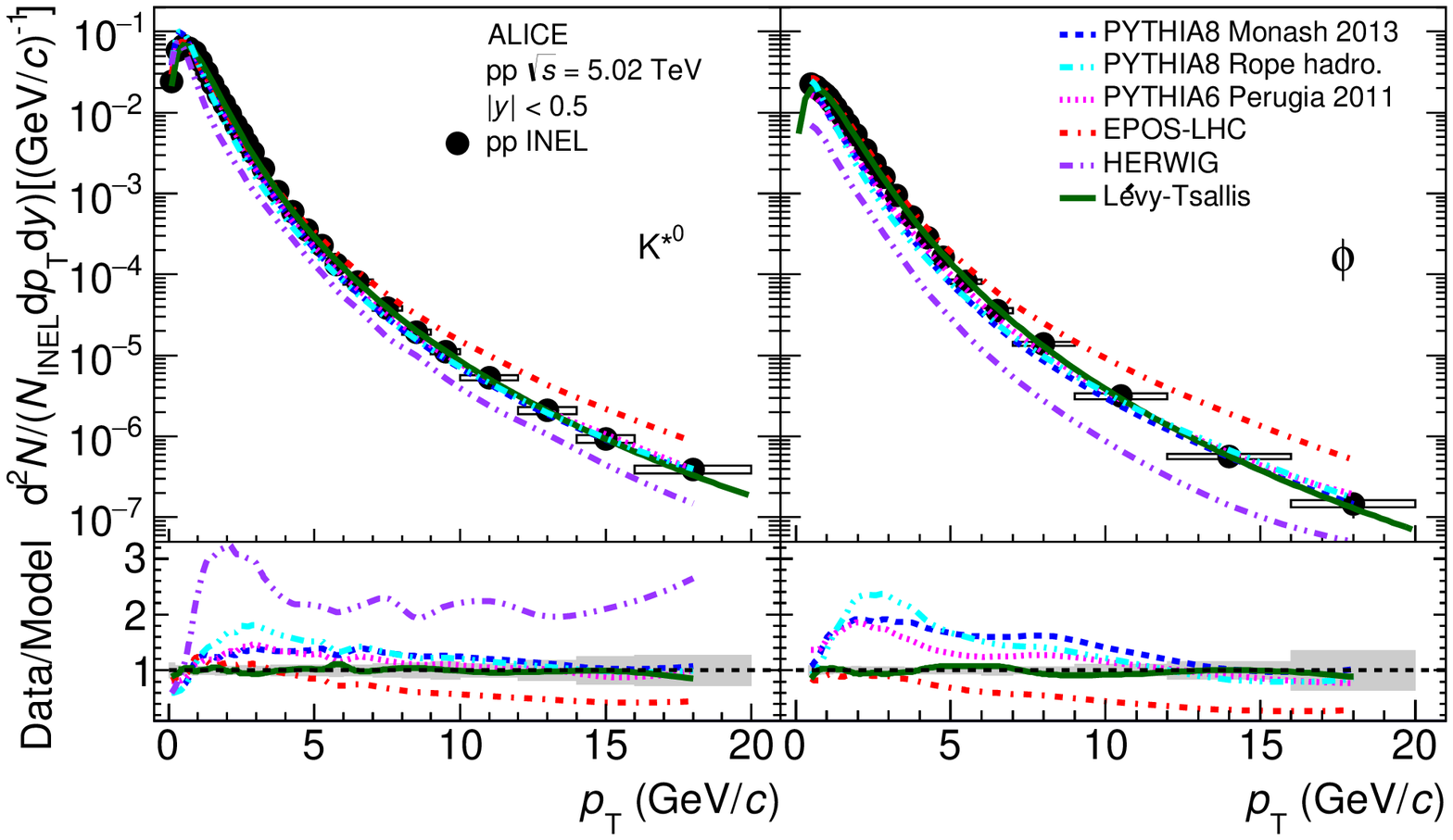}
    \end{center}
    \caption{Transverse momentum spectra of \kstar (left panel) and 
      \phim (right panel) for inelastic \pp collisions at \s~$=$~5.02 
      \TeV. The statistical and systematic uncertainties are shown by 
      bars and boxes, respectively. The results are compared with 
      model calculations from
      PYTHIA 6.4 (Perugia 2011 Tune)~\cite{Sjostrand:2006za,Skands:2010ak}, 
PYTHIA 8.1 (Monash 2013 Tune)~\cite{Sjostrand:2007gs,Skands:2014pea},
PYTHIA 8.2 (Rope hadronization)~\cite{Bierlich:2017sxk},  
      EPOS-LHC~\cite{Pierog:2013ria}, and HERWIG 7.1~\cite{Bahr:2008pv}, which are shown as different 
      dashed lines. The lower panels show the data to model ratio.}
    \label{fig:ptpp}
  \end{figure}
Figure~\ref{fig:ptpp} shows the invariant yields of \kstar and \phim
mesons as a function of \pt for inelastic \pp collisions at \s $=$
5.02 TeV. These first measurements at \s $=$ 5.02 TeV extend to \pt
$=$ 20 GeV$/c$ in the rapidity range of $|y| <$ 0.5. The shape of both
spectra is well described by a L\'{e}vy-Tsallis
function~\cite{Tsallis:1987eu} whose form is given by  
\begin{equation}
\label{eq:spectra_fits:levy}
\frac{\mathrm{d}^2N}{\mathrm{d}p_{\rm T}\rm{d}y} = p_{\rm{T}}\frac{\mathrm{d}N}{\mathrm{d}y}\frac{(n-1)(n-2)}{nC\left[nC+m(n-2)\right]}\left[1+\frac{m_{\rm{T}}-m}{nC}\right]^{-n},
\end{equation}
where $\mathrm{d}N/\mathrm{d}y$, $n$, and $C$ are the parameters of the function that are
determined from the fit to the measured spectra; $m$ is the mass of
the hadron and $m_{\mathrm T}$ is the transverse mass defined as
$\sqrt{p_{\mathrm T}^2 + m^{2}}$.
The L\'{e}vy-Tsallis function provides a fair description of the shape of the transverse momentum spectrum over a wide \pt range, thanks to its two parameters: the exponent $n$ and the inverse slope $C$.
The parameters obtained
from the fit to \kstar and \phim spectra in \pp collisions at \s $=$
5.02 \TeV are given in Table~\ref{table:levypp}. They are similar to the parameters obtained in pp collisions at \s $=$ 7 and 8
\TeV~\cite{Acharya:2019wyb}. The $\chi^{2}/\mathrm{ndf}$ values are less than
unity because the bin-to-bin systematic uncertainties taken in the fit
could be correlated. 
The data are compared to the
corresponding results from the QCD inspired Monte Carlo event
generators like PYTHIA6~\cite{Sjostrand:2006za},
PYTHIA8~\cite{Sjostrand:2007gs}, HERWIG~\cite{Bahr:2008pv}  and
EPOS-LHC~\cite{Pierog:2013ria}.  In PYTHIA model hadronization of
light and heavy quarks is simulated using the Lund string
fragmentation model~\cite{Andersson:1983ia}.
Various PYTHIA tunes have been developed on the basis of
extensive comparisons of Monte Carlo distributions with the minimum bias data from different experiments. Perugia tunes of PYTHIA6 include the revised set of
parameters of fragmentation and flavor which improves the overall
description of the Tevatron data as well as the reliability of the
extrapolations to the LHC measurements~\cite{Skands:2010ak}. Perugia 2011 takes into account the minimum bias and underlying event
data from the LHC at $\sqrt{s}$ $=$ 0.9 and 7 TeV.  The Monash 2013
Tune of PYTHIA8 uses the updated set of hadronization parameters
compared to the previous tunes~\cite{Skands:2014pea}. It gives an overall
good description of kaon data but significantly underestimates the
baryon yields at the LHC. The Rope Hadronization model within the
framework of PYTHIA8 assumes that instead of independent string
fragmentation, the strings overlap to form ropes in the high
multiplicity environment~\cite{Bierlich:2017sxk}.  In the Rope
Hadronization model, the larger and denser collision systems form
color ropes that hadronize with larger string tension leading to
enhanced production of strange hadrons with increasing charged
particle multiplicity. The HERWIG model includes processes such as
coherent parton showers for initial and final state QCD radiation, an
eikonal multiple parton-parton interaction model for the underlying event
and a cluster hadronization model for the  formation of hadrons from
the quarks and gluons produced in the parton
shower~\cite{Bahr:2008pv}.   EPOS-LHC, which is built on the
Parton-Based Gribov Regge Theory, implements a different type of
radial flow for pp collisions, where a very dense system is created in a small volume. 
 The model, utilizing the color exchange mechanism of string
 excitation, is tuned to LHC data~\cite{Pierog:2013ria}. In this
 model, the part of the collision system that has high string or
 parton densities becomes a ``core'' region that may evolve as a
 quark--gluon plasma; this is surrounded by a more dilute ``corona''
 for which fragmentation occurs as in the vacuum. The strangeness
 production is higher in the core region that results in strangeness
 enhancement with increasing multiplicity.  

For \kstar, all the tunes of
PYTHIA model overestimate the data for \pt $<$ 0.5 GeV$/c$,
underestimate the data  in the intermediate \pt region and give a
better description for \pt $>$ 10 GeV$/c$. For \phim,  all the tunes
of PYTHIA model give a better description of the data for \pt $>$ 10
GeV$/c$ and underestimate the data for lower \pt region. The
deviations in limited \pt ranges observed between data and PYTHIA for
both \kstar and \phim are similar to those reported for \s $=$
2.76~\cite{Adam:2017zbf} and 7 TeV~\cite{Abelev:2012hy}. The EPOS-LHC
model results are in agreement with the data for \pt $<$ 5 GeV$/c$
(3.5 GeV$/c$) and overestimate the data at higher \pt for \kstar
(\phim). HERWIG does not describe the data for both \kstar and \phim
over the measured \pt region. 

\setlength{\tabcolsep}{19pt}
\begin{table*}
\caption{L\'{e}vy-Tsallis fit parameters for \kstar and \phim meson \pt 
 spectra in \pp collisions at \s $=$ 5.02 \TeV. The errors on the fit
 parameters result from the total (quadrature sum of statistical and
 systematic) uncertainties on the data.}
\begin{tabular}{ c c c c c}
\hline\hline 
Resonance & $n$ & $C$ (GeV) & $\chi^{2}/$ndf\\\hline 
\kstar & $7.05 \pm  0.14$ & $0.267 \pm 0.006$ &  0.23
  \\\hline 
\phim & $7.47 \pm  0.16$ & $0.309 \pm 0.006$ &  0.62
  \\\hline \hline 
\end{tabular}
\label{table:levypp}
\end{table*}

\begin{figure}[tb]
    \begin{center}
\includegraphics[scale = 0.7]{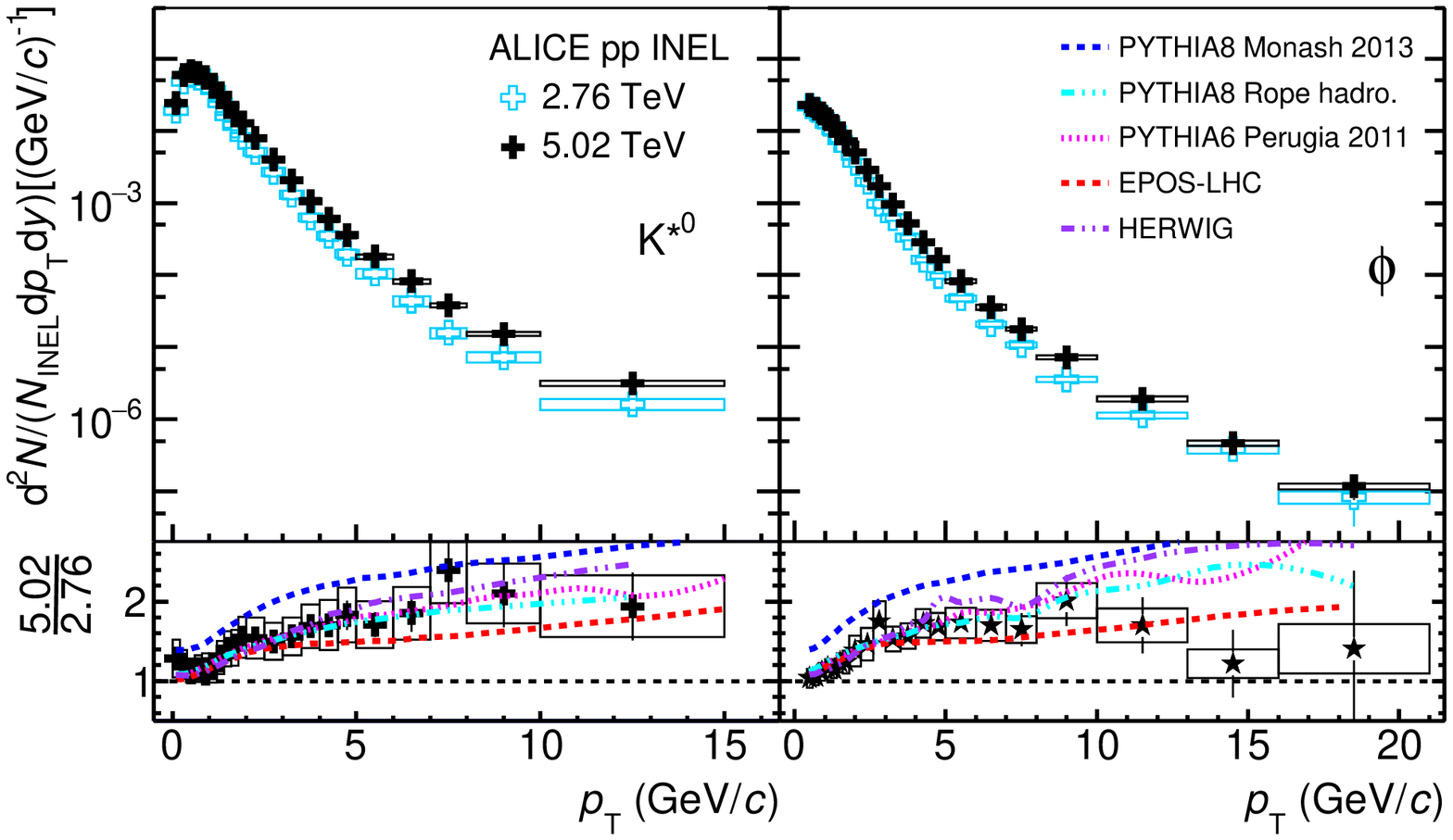}
    \end{center}
    \caption{Comparison of transverse momentum spectra of \kstar (left panel) and 
      \phim (right panel) for inelastic \pp collisions at \s $=$  5.02 
      \TeV (solid markers) and 2.76 \TeV (open markers) ~\cite{Adam:2017zbf} . The lower panels show the ratio of the \pt
      spectra at \s $=$ 5.02 \TeV to those from \s $=$ 2.76 \TeV. The
      statistical uncertainties on the data are shown by bars and the
      systematic uncertainties by boxes. The ratios are compared with 
      model calculations from
      the PYTHIA 6.4 (Perugia 2011 Tune)~\cite{Sjostrand:2006za,Skands:2010ak}, 
PYTHIA 8.1 (Monash 2013 Tune)~\cite{Sjostrand:2007gs,Skands:2014pea},
PYTHIA 8.2 (Rope hadronization)~\cite{Bierlich:2017sxk},  
       EPOS-LHC~\cite{Pierog:2013ria}, and HERWIG 7.1~\cite{Bahr:2008pv}, which are shown as different
      dashed lines.}
    \label{fig:ptppen}
  \end{figure}
Figure~\ref{fig:ptppen} shows the comparison of \pt spectra of \kstar
and \phim mesons  between \s $=$ 5.02 and 2.76 \TeV. The yields of
both \kstar and \phim mesons are higher at \s $=$ 5.02 \TeV compared
to \s $=$ 2.76 \TeV.  The ratio of the \pt spectra at \s $=$ 5.02 to
2.76 \TeV as a function of \pt shows that the differential yield ratio
increases with \pt for both \kstar and \phim mesons and show a hint of
saturation at higher \pt. These results further help in understanding
the nuclear modification factor (derived using pp reference spectra)
for \PbPb collisions that will be discussed in Sec~\ref{sec:RAA}. The
ratios are compared to the corresponding calculations from the PYTHIA6
(Perugia 2011 Tune), PYTHIA8 (Monash 2013 Tune), PYTHIA8 (Rope hadronization),  
EPOS-LHC, and HERWIG models. For the \kstar meson, all the models
except PYTHIA8-Monash 2013 Tune are in good agreement with the
measurements within  uncertainties for the whole \pt range.  For the
\phim meson, all the models except PYTHIA8-Monash 2013 Tune are in
good agreement with the measurements within uncertainties for \pt $<$
8 GeV$/c$.  The ratio of the \pt spectra for \kstar and \phim from the Monash
2013 Tune of PYTHIA8 are higher than the measurements for all \pt.
All the models presented here fail quantitatively and/or qualitatively to describe the \kstar and \phim data over the entire measured \pt range. It has proven challenging for event generators to accurately model the \pt distributions of such resonances~\cite{Abelev:2012hy,Acharya:2019wyb,Acharya:2020uxl,Adam:2017zbf}. Thus, the data and model comparisons may provide valuable inputs to tune the MC event generators so that one get a unified physics description of present results.

\subsection{\pt spectra in \PbPb collisions}\label{sec:pTspectPbPb}
\begin{figure}[tb]
    \begin{center}
      \includegraphics[scale = 0.7]{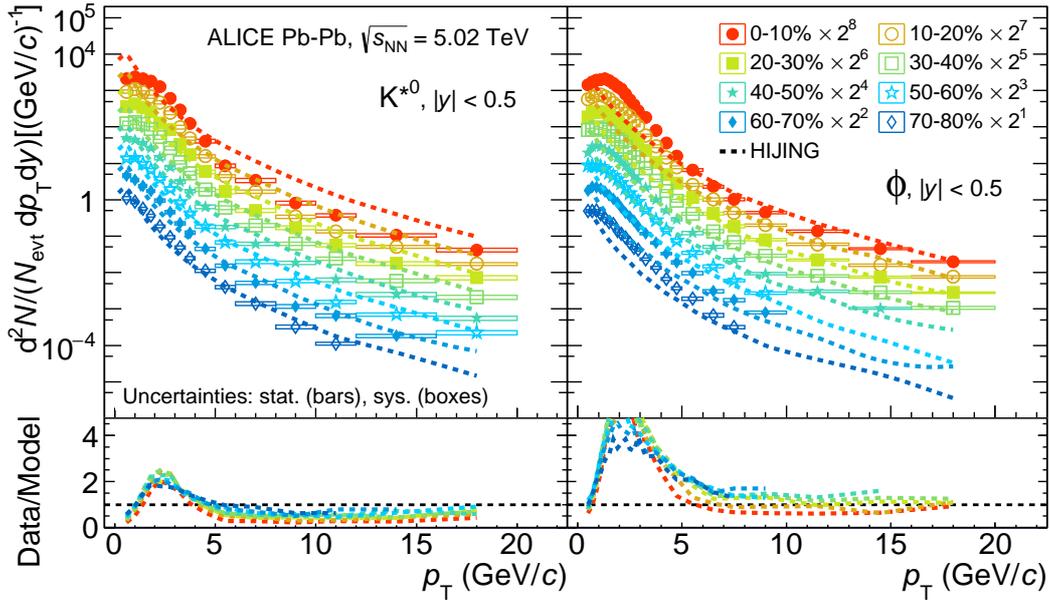} 
    \end{center}
    \caption{Transverse momentum spectra of \kstar (left panel) and
      \phim (right panel) at midrapidity in various centrality
      classes in \PbPb collisions at \snn $=$  5.02 \TeV. The
      statistical uncertainties on the data are shown by bars and the
      systematic uncertainties by boxes. The results are compared with 
      model calculations from
      HIJING 1.36~\cite{Wang:1991hta}, which are shown as dashed lines. The 
      lower panel shows the data to model ratios. 
    } \label{fig:ptPbPb}
  \end{figure}
Figure~\ref{fig:ptPbPb} shows the \pt distributions for \kstar and
\phim mesons in \PbPb collisions  at \snn $=$ 5.02 TeV. The
measurements are carried out at midrapidity ($|y| <$ 0.5) for eight
different centrality classes. For the most central collisions (0--10\%), the measurements
extend up to \pt $=$ 20 GeV$/c$ for both \kstar and \phim mesons. The HIJING model was used to calculate the reconstruction efficiency of \kstar and
\phim mesons in \PbPb collisions, hence the
data are compared to the corresponding results from the HIJING model~\cite{Wang:1991hta}. For \kstar meson, HIJING does not describe
the data for all centrality classes over the whole measured \pt
region. For \phim meson, the model gives a good agreement with the
data in mid-central collisions for \pt~$>$~7~GeV$/c$, however does not
describe the data for most of the centrality classes. The \pt spectra
are further characterized by the integrated yields (d$N/$d$y$) and the
average transverse momentum ($\langle p_{\mathrm T} \rangle$) which
are discussed in the next subsection. 

\begin{figure}[tb]
    \begin{center}
 \includegraphics[scale = 0.7]{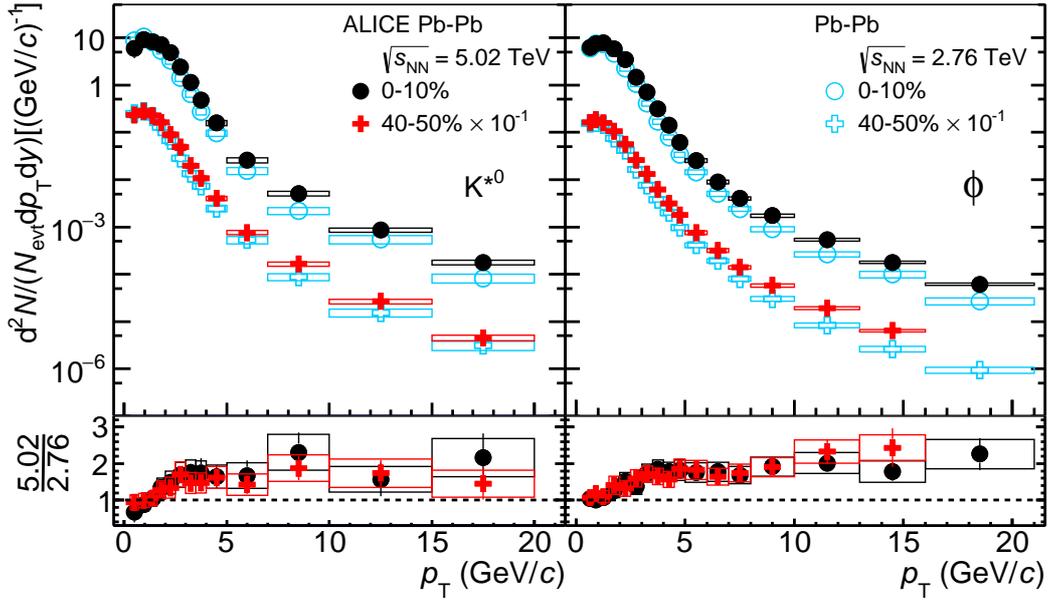}
    \end{center}
    \caption{Comparison of transverse momentum spectra of \kstar (left
      panel) and \phim (right panel) in central (0--10\%) and
      mid-central (40--50\%) centrality classes in \PbPb collisions at
      \snn $=$  5.02  (solid marker) and 2.76 (open marker) \TeV~\cite{Adam:2017zbf,Abelev:2014uua}. The lower panels show the ratio of the \pt
      spectra at \snn~$=$~5.02~\TeV to those from \snn~$=$~2.76~\TeV. The
      statistical uncertainties on the data are shown by bars and the
      systematic uncertainties by boxes.} 
    \label{fig:ptPbPben}
  \end{figure}
Figure~\ref{fig:ptPbPben} shows the comparison of the \pt spectra for
\kstar and \phim mesons in \PbPb collisions for the 0--10\% and 40--50\%
centrality classes at \snn $=$ 5.02 and 2.76 TeV. The ratios of the
\pt spectra increase with \pt and then tend to saturate at high \pt
for both mesons in central as well as in peripheral collisions, as
also observed in \pp collisions  (Fig.~\ref{fig:ptppen}). These
results are useful in understanding the energy dependence of the
nuclear modification factor which is discussed in
Sec~\ref{sec:RAA}. For $p_{\mathrm{T}}$ $>$ 5.0 GeV$/c$, the
$p_{\mathrm{T}}$ differential yields at 5.02 TeV are $\approx$1.8 times
higher than those measured at 2.76 TeV.  

\begin{figure}[tb]
    \begin{center}
    \includegraphics[width = 0.45\textwidth]{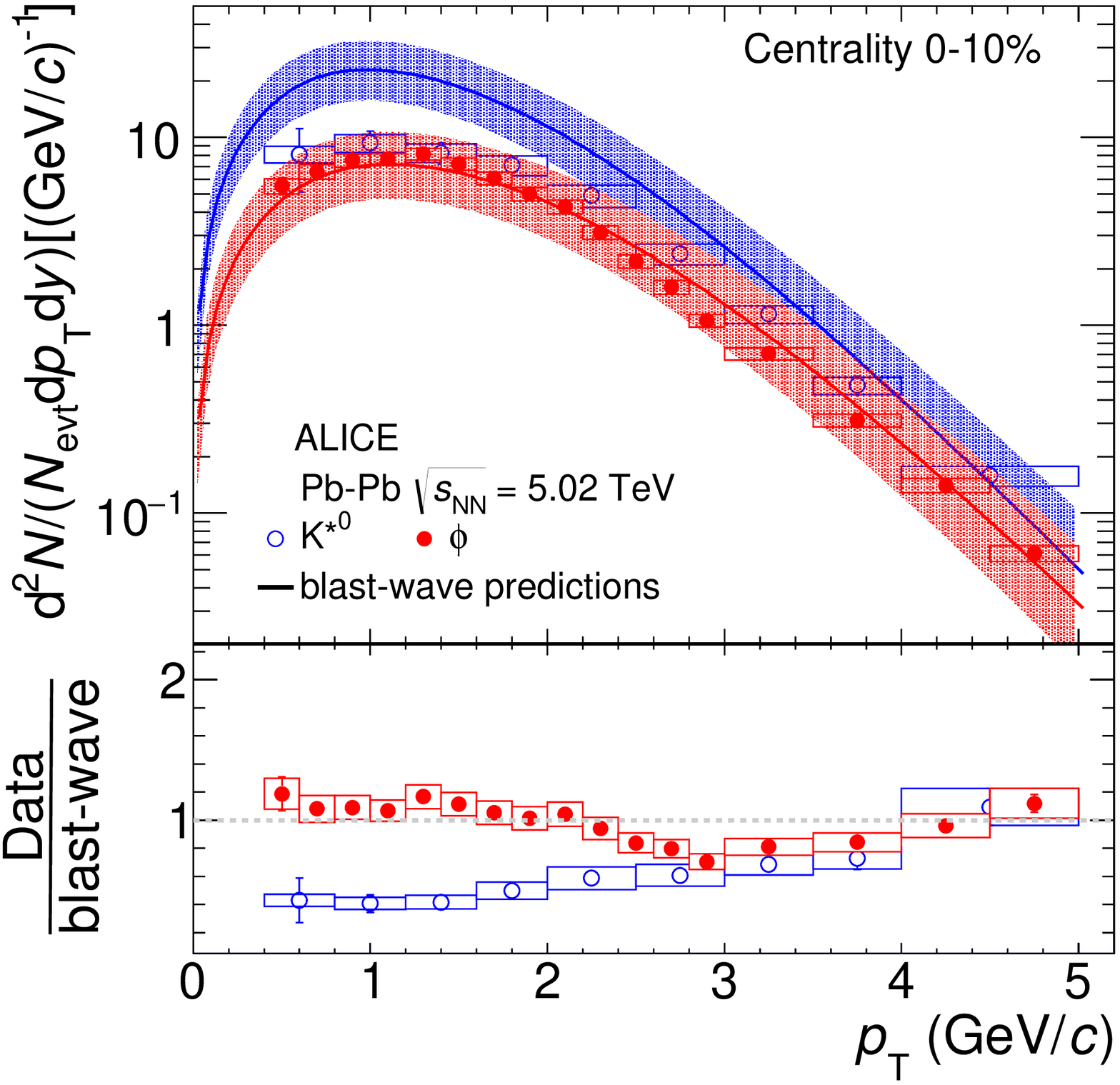}
    \includegraphics[width = 0.45\textwidth]{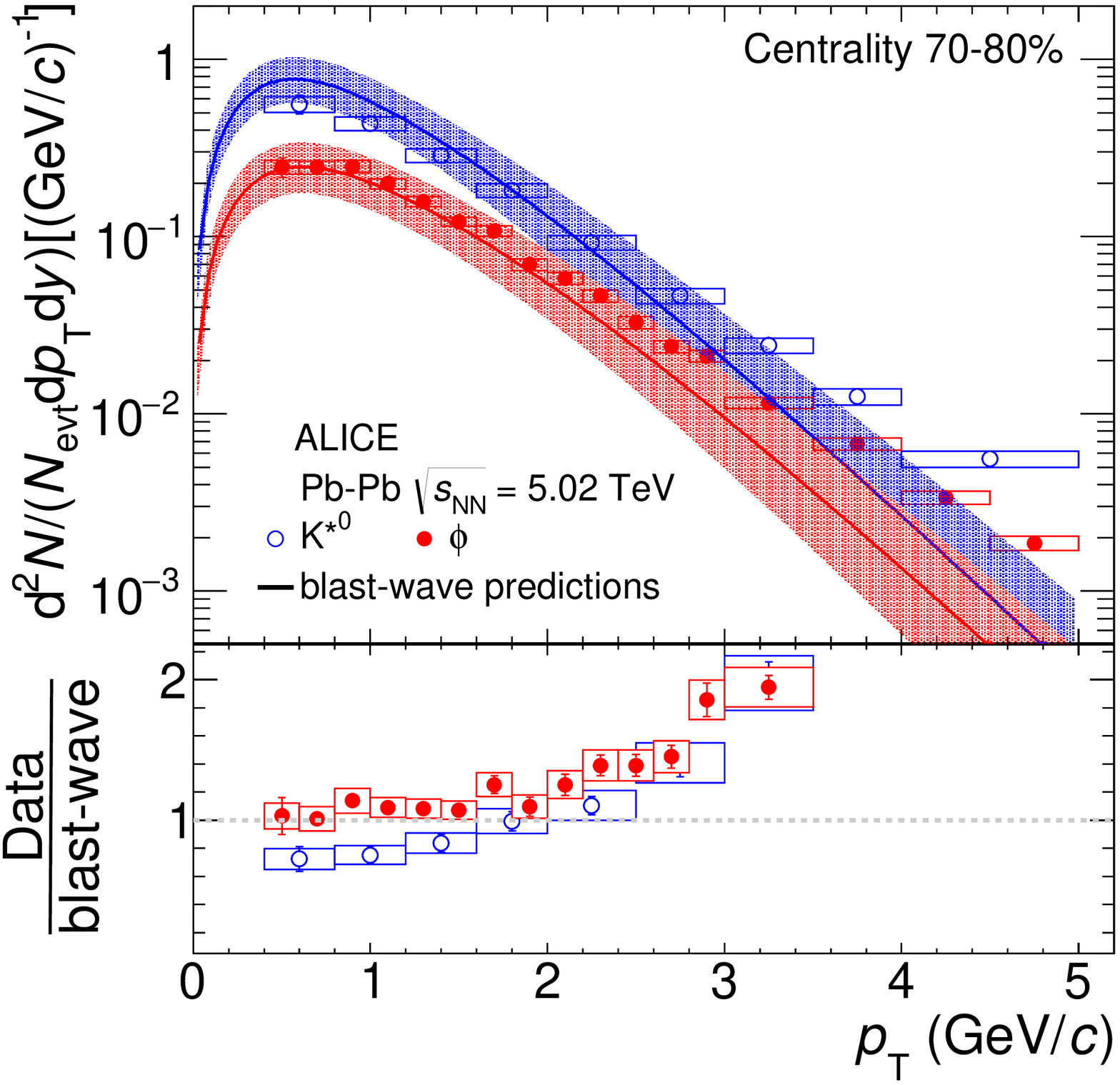}	
    \end{center}
    \caption{Transverse-momentum distributions of \kstar and \phim
      resonances in Pb--Pb collisions at \snn $=$ 5.02 \TeV along with
      expected distributions for central (left) and peripheral (right)
      collisions. The shapes of the expected distributions are given
      by Boltzmann-Gibbs blast-wave
      functions~\cite{Schnedermann:1993ws} using parameters obtained
      from fits to pion, kaon, and (anti)proton \pt distributions
     ~\cite{Acharya:2019yoi}  (see text for details).  The shaded
      bands indicate the uncertainties in the fit parameters of the
      model distributions.  The lower panels show the ratios of the
      measured distribution to the values from the model. 
   }
    \label{fig:PbPbBW}
\end{figure}

A blast-wave model, which does not include rescattering effects, is
used to investigate the \pt dependence of resonance yield suppression. 
Previously, in \PbPb collisions at \snn $=$ 2.76 \TeV~\cite{Abelev:2014uua}, 
the \kstar and \phim \pt spectra were compared to the expected
distributions based on the blast-wave model~\cite{Schnedermann:1993ws}
 using parameters obtained from the combined fit to $\pi^{\pm}$,
 K$^{\pm}$, and p($\bar{\rm p}$) spectra.  A suppression of the \kstar
 yield with respect to the blast-wave distribution was observed for
 \pt~$<$~3~GeV$/c$ in central collisions. This suppression is attributed
 to rescattering of resonance decay products in the hadronic phase
 that reduces the measurable yield of \kstar
 mesons~\cite{Acharya:2019qge}.  The lack of similar suppression for
 the \phim meson is interpreted as being due to the absence of
 rescattering, as it mostly decays outside the fireball because of its
 longer lifetime.  We have  carried out a similar exercise for \PbPb
 collisions at \snn $=$ 5.02 TeV. The Boltzmann-Gibbs blast-wave
 function is a three parameter simplified hydrodynamic model, which
 assumes that the emitted particles are locally thermalized in a
 uniform-density source at a kinetic freezeout temperature
 $T_{\mathrm{kin}}$ and move with a common collective transverse
 radial flow velocity field. It is given by~\cite{Schnedermann:1993ws} 
\begin{equation}
\label{eq:results:spectra:blastwave}
\frac{1}{\pt}\frac{\dd N}{\dd\pt}\propto \int_{0}^{R}r\;\dd r\;\mT\; I_{0}\negthickspace\left(\frac{\pt\;\mathrm{sinh}\rho}{T_{\mathrm{kin}}}\right)K_{1}\negthickspace\left(\frac{\mT\;\mathrm{cosh}\rho}{T_{\mathrm{kin}}}\right),
\end{equation}

where $I_{0}$ and $K_{1}$ are the modified Bessel functions, $R$ is
the fireball radius, and $r$ is the  radial distance in the transverse
plane.  The velocity profile $\rho$ is defined as: 

\begin{equation}
\label{eq:results:spectra:rho}
\rho=\mathrm{tanh}^{-1}\beta_{\mathrm{T}}(\mathrm{r})=\mathrm{tanh}^{-1}\left[\left(\frac{r}{R}\right)^{n}\beta_{\mathrm{s}}\right],
\end{equation}

where $\beta_{\mathrm{T}}(\mathrm{r})$ is the transverse expansion velocity and 
$\beta_{\mathrm{s}}$ is the transverse expansion velocity at the surface.  The free 
parameters in the fits are $T_{\mathrm{kin}}$, $\beta_{\mathrm{s}}$, and the velocity 
profile exponent $n$.
For the current study, the above parameters, listed in Table~\ref{table:bw}, are fixed to the values from fits to the charged pion, kaon, and (anti)proton  \pt distributions in \PbPb collisions at \snn $=$ 5.02 \TeV~\cite{Acharya:2019yoi}.

\setlength{\tabcolsep}{29pt}
\begin{table*}
\caption{Blast-wave parameters from fit to charged pion, kaon, and 
  proton spectra in \PbPb collisions at \snn $=$ 5.02 \TeV.}
\begin{tabular}{ c c c c }
\hline\hline 
Centrality & $T_{\mathrm{kin}}$ (\GeV) & $\langle \beta_{\mathrm{T}} \rangle$ & $n$ \\\hline 
0--10\% & $0.091 \pm 0.003$  & $0.662 \pm 0.003 $ & $0.735 \pm 0.013$  \\\hline 
70--80\% & $0.147 \pm 0.006$  & $0.435 \pm 0.011 $ & $1.678 \pm 0.088$   \\\hline \hline 
\end{tabular}
\label{table:bw}
\end{table*}

Figure~\ref{fig:PbPbBW} shows the expected \kstar and \phim \pt
distributions from the  blast-wave model (as solid lines), the
measured  resonance \pt distributions, and the ratios of the
measurement to the blast-wave model for central (0--10\%) and
peripheral (70--80\%) collisions. The expected distributions are
normalized so that their integrals are equal to the measured yield of
charged kaons in Pb--Pb collisions at \snn $=$ 5.02
\TeV~\cite{Acharya:2019yoi} multiplied by the $\mathrm{K^{*0}/K}$ and
$\mathrm{\phi/K}$ ratios from pp collisions at \s $=$ 5.02 \TeV. A
similar procedure of normalization was used in \PbPb collisions at
\snn $=$ 2.76 TeV~\cite{Abelev:2014uua}. The ratio for the \phim meson
\pt distribution is close to unity and no significant differences are
observed in central or peripheral collisions for $\pt<2$~GeV$/c$.  On
the other hand, the data$/$blast-wave ratio for the \kstar is lower than unity with a deviation of 40--60\% for $\pt<3$~GeV$/c$ in central collisions. 
In peripheral collisions, the  data$/$blast-wave ratio for the \kstar
shows a significantly smaller deviation from unity for $\pt<2$~GeV$/c$
relative to central collisions. Both \kstar and \phim show a similar
deviation for \pt $>$ 3 GeV$/c$ ($>$ 2.5 GeV$/c$) in central
(peripheral) collisions. The blast-wave model is expected to describe
the measured \pt distributions over the entire \pt range if these are
driven purely by the collective radial expansion of the system. The
model describes the  data over a wider \pt interval for central \PbPb
collisions than for peripheral collisions as observed for $\pi$, K, p
in \PbPb collisions at \snn $=$ 5.02 TeV~\cite{Acharya:2019yoi}. 
For \PbPb collisions, the average transverse velocity $\langle
\beta_{\mathrm{T}}\rangle$ is observed  to increase with centrality
while $T_{\mathrm{kin}}$ decreases~\cite{Acharya:2019yoi}. For central
\PbPb collisions, the shape of the \pt distributions of \kstar and
\phim mesons for  $\pt<2$~GeV$/c$ are consistent with the blast-wave
parameterization within uncertainties. The suppression of yields of
\kstar with respect to the blast-wave model expectation in central
collisions, relative to peripheral collisions and \phim mesons, is
consistent with the dominance of  rescattering effects in the medium
formed in \PbPb collisions at \snn $=$ 5.02 TeV.

\subsection{d$N/$d$y$ and $\langle p_{\mathrm T} \rangle$}
\begin{figure}[tb]
    \begin{center}
      \includegraphics[scale = 0.7]{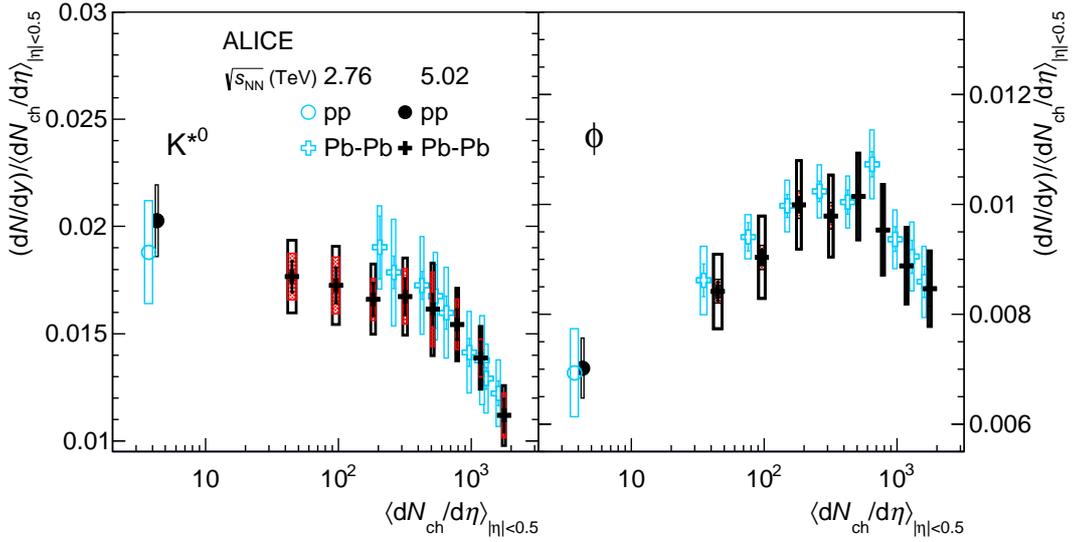}	
    \end{center}
\caption{Transverse momentum integrated yield (d$N/$d$y$) of \kstar
  and \phim as a function of the averaged charged particle density 
  (\avdndeta$_{|\eta|\textless0.5}$) in pp and \PbPb collisions at 
  \snn $=$ 5.02 \TeV and compared with result from Pb--Pb collisions at   
  \snn $=$ 2.76 \TeV~\cite{Adam:2017zbf,Abelev:2014uua}. The
  statistical  uncertainties on the data 
  are shown by bars and the systematic uncertainties by boxes. The
  shaded box shows the uncorrelated systematic uncertainties.} 
\label{fig:dNdyPbPb}
\end{figure}
Figure~\ref{fig:dNdyPbPb} shows the \pt-integrated yield d$N/$d$y$ of
\kstar and \phim mesons scaled by the average charged particle multiplicity 
measured at midrapidity (\avdndeta$_{|\eta|\textless0.5}$) as a
function of \avdndeta$_{|\eta|\textless0.5}$ for \PbPb and \pp
collisions at \snn $=$ 5.02 \TeV. The \pt-integrated yields
(d$N/$d$y$) have been obtained by integrating the spectra over \pt
using the measured data and a blast-wave function (L\'{e}vy-Tsallis
function) in the unmeasured regions for \PbPb (\pp) collisions. 
The fraction of the yields from the extrapolation to 
the total for \kstar(\phim) mesons is 0.09 (0.08) in the 0--10\%
centrality class, and is 0.16 (0.12) in the 70--80\% centrality class. 
This fraction is 0.17 for $\phim$ in pp, whereas no extrapolation is
needed for \kstar. For comparison, the corresponding results from \snn
$=$ 2.76 TeV  are also shown in Fig.~\ref{fig:dNdyPbPb}.
The dependence of the normalized d$N/$d$y$ on \avdndeta~is found to be
the same regardless of the beam energy. 

\begin{figure}[tb]
    \begin{center}
      \includegraphics[scale = 0.7]{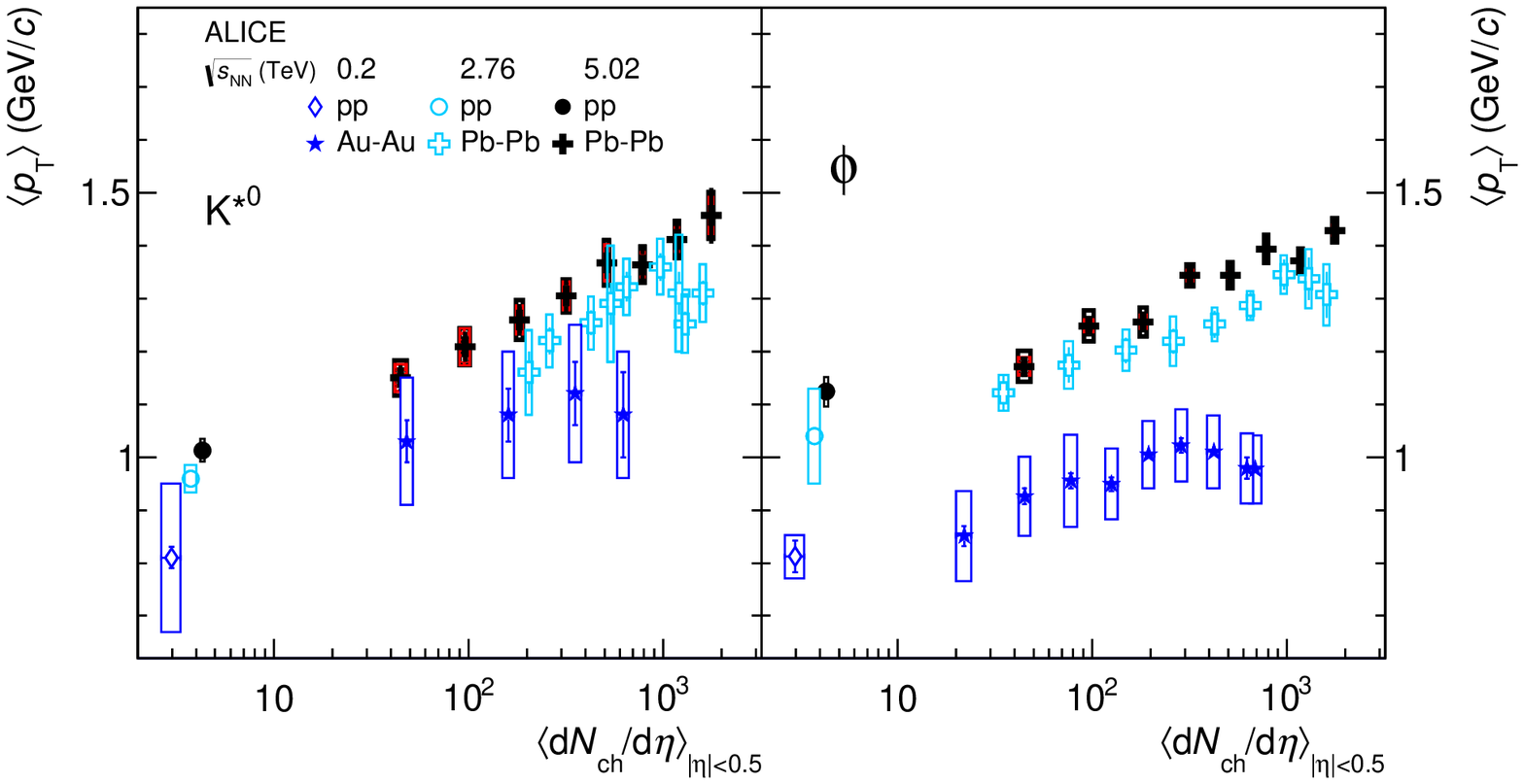}
    \end{center}
\caption{Mean transverse momentum of \kstar and \phim as a  function of the 
averaged charged particle density (\avdndeta$_{|\eta|\textless0.5}$)
in \PbPb and pp collisions at \snn $=$ 5.02 \TeV. Also shown for
comparison are the corresponding values from  \PbPb and pp collisions
at \snn $=$ 2.76 \TeV~\cite{Adam:2017zbf,Abelev:2014uua}, and Au--Au
and pp collisions at \snn $=$ 200
\GeV~\cite{Adams:2004ep,Adler:2002sw,Abelev:2008zk,Abelev:2008aa}. The
statistical  uncertainties on the data are shown by bars and the
systematic uncertainties by boxes.  The shaded box shows the
uncorrelated systematic uncertainties. 
} 
    \label{fig:meanpt}
\end{figure}

The average transverse momentum \meanpt values for the \kstar and
\phim resonances are  obtained by using the data in the measured
region and a blast-wave function (L\'{e}vy-Tsallis function) in the
unmeasured regions for \PbPb (pp) collisions. Figure~\ref{fig:meanpt}
shows \meanpt values obtained at midrapidity ($|y| <$ 0.5) as a
function of \avdndeta$_{|\eta|\textless0.5}$ for \PbPb and \pp
collisions at \snn $=$ 5.02 \TeV. The \meanpt values increase with
charged particle  multiplicity.  The \meanpt of \kstar and \phim
mesons, which have similar masses, are similar for events with the
same \avdndeta$_{|\eta|\textless0.5}$ in \PbPb collisions. Both of
these features  are consistent with the picture of a growing
contribution of radial flow with increasing
\avdndeta$_{|\eta|\textless0.5}$ for  the system formed in \PbPb
collisions at \snn $=$ 5.02 \TeV~\cite{Acharya:2019yoi}.  
These results are also compared to the corresponding results from
\PbPb collisions at \snn $=$ 2.76
\TeV~\cite{Adam:2017zbf,Abelev:2014uua}, and to Au--Au and pp
collisions at \snn $=$ 200
\GeV~\cite{Adams:2004ep,Adler:2002sw,Abelev:2008zk,Abelev:2008aa}. The
\meanpt values are larger for higher energy collisions at similar
values of \avdndeta$_{|\eta|\textless0.5}$, although the uncertainties
are also larger at lower collision energies for \kstar. The
qualitative features of the dependence of \kstar and \phim meson
\meanpt on \avdndeta$_{|\eta|\textless0.5}$ are similar at RHIC and
LHC energies. The values of d$N/$d$y$ and \meanpt of \kstar and \phim
measured in Pb--Pb and pp collisions at \snn~$=$~5.02 TeV are given in
Table~\ref{table:dndymeanpt}. 

\setlength{\tabcolsep}{19pt}
\begin{table*}
\caption{The values of d$N/$d$y$ and \meanpt measured in Pb--Pb and pp
collisions at \snn~$=$~5.02 TeV. Pb--Pb results are shown for the
different centrality classes. In each entry the first uncertainty is
statistical and the second one is systematic.}
\begin{tabular}{ c c c }
\hline\hline 
 &  \kstar &   \\
\hline 
Centrality class & d$N/$d$y$  & \meanpt \\
\hline 
0--10\% & $19.726~\pm~1.357~\pm~2.475$ & $1.457~\pm~0.053~\pm~0.044$ \\
10--20\% & $16.363~\pm~0.895~\pm~1.725$ & $1.411~\pm~0.039~\pm~0.035$ \\
20--30\% & $12.129~\pm~0.547~\pm~1.331$& $1.364~\pm~0.031~\pm~0.034$ \\
30--40\% & $8.260~\pm~0.361~\pm~1.111$& $1.368~\pm~0.030~\pm~0.043$ \\
40--50\% & $5.321~\pm~0.268~\pm~0.574$& $1.305~\pm~0.034~\pm~0.032$ \\
50--60\% & $3.039~\pm~0.140~\pm~0.300$& $1.259~\pm~0.030~\pm~0.037$ \\
60--70\% & $1.661~\pm~0.079~\pm~0.176$& $1.209~\pm~0.028~\pm~0.035$ \\
70--80\% & $0.793~\pm~0.033~\pm~0.076$ & $1.150~\pm~0.024~\pm~0.033$ \\
\hline
 &  \phim &   \\
\hline 
Centrality class & d$N/$d$y$  & \meanpt \\
\hline 
0--10\% &$14.937~\pm~0.154~\pm~1.210$ & $1.429~\pm~0.008~\pm~0.023$ \\
10--20\% &$10.483~\pm~0.113~\pm~0.828$ & $1.371~\pm~0.008~\pm~0.024$ \\
20--30\% &$7.497~\pm~0.083~\pm~0.647$ & $1.393~\pm~0.009~\pm~0.027$ \\
30--40\% &$5.192~\pm~0.056~\pm~0.404$ & $1.344~\pm~0.008~\pm~0.024$ \\
40--50\% &$3.113~\pm~0.038~\pm~0.237$ & $1.343~\pm~0.009~\pm~0.021$ \\
50--60\% &$1.829~\pm~0.024~\pm~0.148$ & $1.256~\pm~0.010~\pm~0.027$ \\
60--70\% &$0.870~\pm~0.013~\pm~0.072$ & $1.248~\pm~0.011~\pm~0.029$ \\
70--80\% &$0.378~\pm~0.008~\pm~0.031$  & $1.171~\pm~0.014~\pm~0.029$ \\
 \hline
&  pp collisions &   \\
\hline 
Particle specie & d$N/$d$y$  & \meanpt  \\
\hline 
\kstar & $0.0872~\pm~0.0006~\pm~0.0072$  &$1.0131~\pm~0.0038~\pm~0.0216$ \\
\phim & $0.0302~\pm~0.0002~\pm~0.0024$ &$1.1239~\pm~0.0047~\pm~0.0282$ \\
\hline\hline
\end{tabular}
\label{table:dndymeanpt}
\end{table*}

\begin{figure}[tb]
    \begin{center}
    \includegraphics[scale = 0.5]{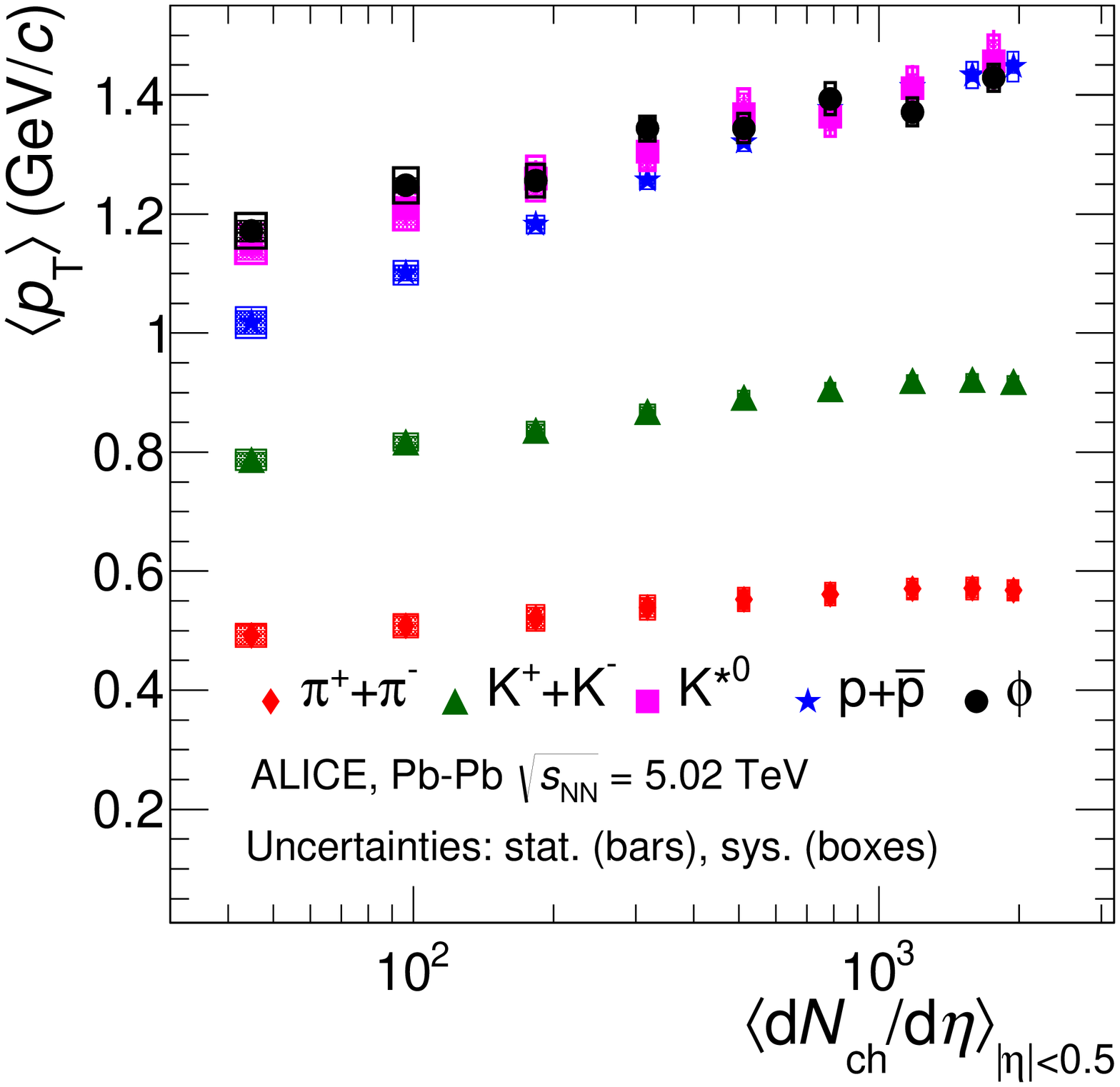}
    \end{center}
\caption{Mean transverse momentum of \kstar and \phim as a 
  function of the averaged charged particle density 
  (\avdndeta$_{|\eta|\textless0.5}$) in \PbPb collisions at \snn $=$
  5.02 \TeV compared with the corresponding values for other
  identified particles such as $\pi^{\pm}$, K$^{\pm}$, and p($\bar
  {\rm p}$)~\cite{Acharya:2019yoi}. The statistical uncertainties on
  the data are shown by bars and the systematic uncertainties by
  boxes. The shaded boxes show the uncorrelated systematic
  uncertainties.} \label{fig:meanptpar}
\end{figure}

Figure~\ref{fig:meanptpar} compares the \meanpt of \kstar and
\phim  as a function of \avdndeta$_{|\eta|\textless0.5}$ with the
respective values for $\pi^{\pm}$, K$^{\pm}$, and p($\bar {\rm
  p}$)~\cite{Acharya:2019yoi} for \PbPb collisions at \snn $=$ 5.02
TeV. All the hadrons exhibit an increase in \meanpt from peripheral to
central \PbPb collisions: the largest increase is observed for protons, followed by
the \kstar and \phim mesons, and then by K and $\pi$.  
The rise in the \meanpt values is steeper for hadrons with higher
mass, as expected in presence of a radial flow effect. For
\avdndeta$_{|\eta|\textless0.5}$ $>$ 300, the \meanpt values of
\kstar, p, and \phim hadrons follow a similar trend and have quantitatively  
similar values within uncertainties at a given
\avdndeta$_{|\eta|\textless0.5}$ value. The masses of these hadrons
are similar, \kstar $\approx$ 896 MeV$/c^{2}$, p $\approx$ 938 MeV$/c^{2}$,
and \phim $\approx$ 1019 MeV$/c^{2}$. The hadron mass dependence of
\meanpt is  consistent with the expectation from a hydrodynamic
evolution of the  system formed in \PbPb collisions at
\snn~$=$~5.02~\TeV for \avdndeta$_{|\eta|\textless0.5}$ $>$ 300.  In
peripheral collisions (\avdndeta$_{|\eta|\textless0.5}$ $<$ 300),
\meanpt of proton is lower than those of \kstar and \phim, indicating
the breaking of mass ordering while going towards peripheral Pb--Pb
collisions. 

\subsection{Particle ratios}\label{sec_partRatio}
\begin{figure}[tb]
    \begin{center}
\includegraphics[scale = 0.7]{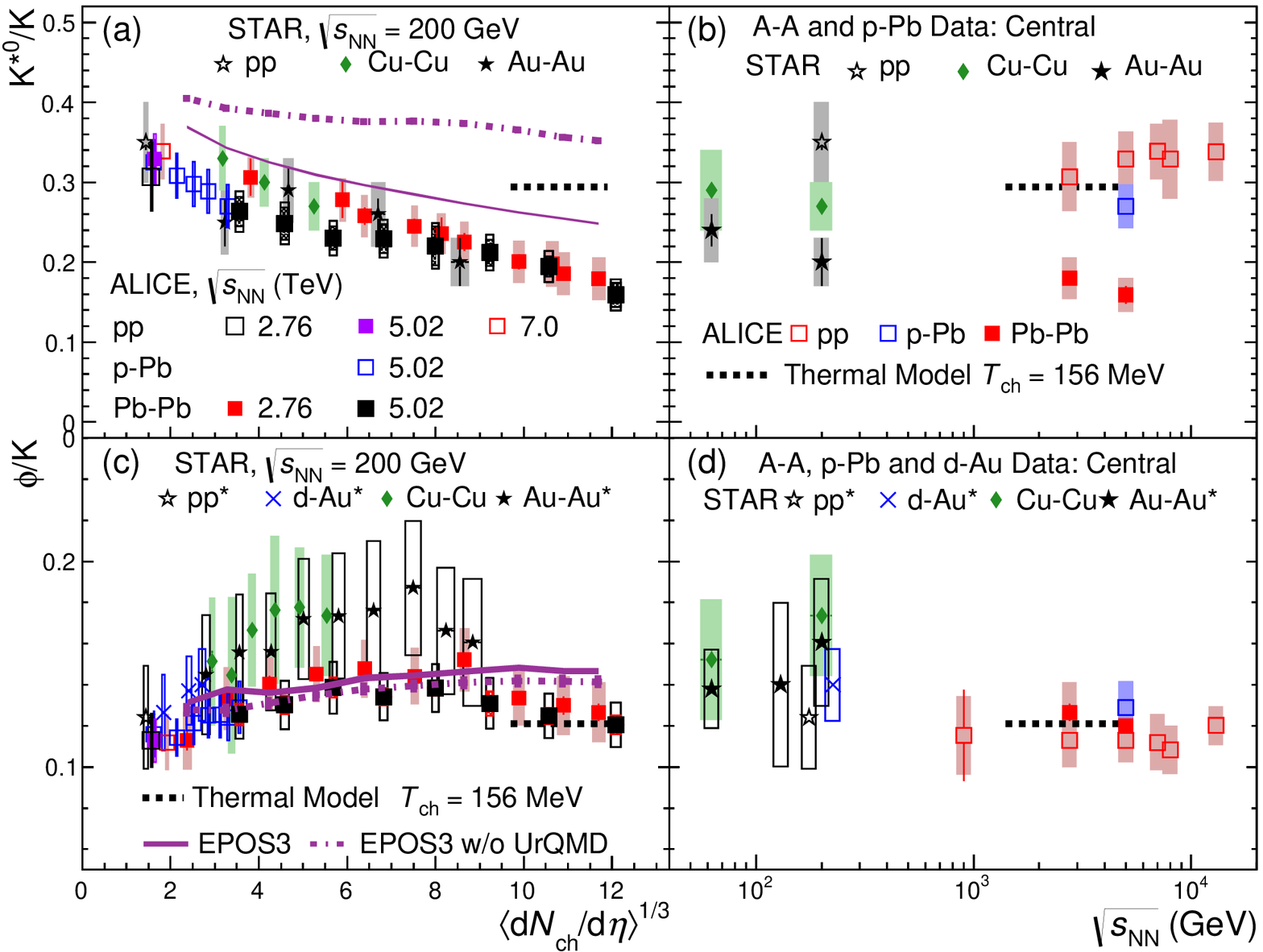}
    \end{center}
    \caption{(Color online)  Particle ratios $\mathrm{K^{*0}/K}$
      [panels (a) and (b)] and $\mathrm{\phi/K}$ [panels (c) and (d)]
      in \pp, \dAu, \pPb, \CuCu, \AuAu, and \PbPb
      collisions~\cite{Adam:2017zbf,Abelev:2014uua,Aggarwal:2010mt,Adams:2004ep,Adler:2002sw,Abelev:2008zk,Abelev:2008aa,Abelev:2008yz,Adam:2016bpr,Acharya:2018orn,Abelev:2012hy}. In
      panels (a) and (c) these ratios are presented as a function of
      \avdndeta$^{1/3}$. These ratios are measured as \kskm~and
      \phikm~in \PbPb collisions at \snn $=$ 2.76 TeV and in STAR experiment. The values of
      \avdndeta$_{|\eta|\textless0.5}$ were measured at midrapidity.
      In panels (b) and (d), these ratios are presented for \pp,
      central \dAu, \pPb, and heavy-ion collisions as a function of
      \snn.  The values given by a grand-canonical thermal model with
      a chemical freezeout temperature of 156~MeV are also
      shown~\cite{Stachel:2013zma}.  For quantities marked ``*", boxes
      represent the total uncertainty (separate uncertainties are not
      reported).  Otherwise, bars represent the statistical
      uncertainties and boxes represent the systematic uncertainties
      (including centrality-uncorrelated and centrality-correlated
      components). EPOS3 model predictions~\cite{Knospe:2015nva} of
      $\mathrm{K^{*0}/K}$ and $\mathrm{\phi/K}$ ratios in \PbPb
      collisions are also shown as violet lines.} \label{fig:kstarphikratio}
  \end{figure}
Figure~\ref{fig:kstarphikratio} shows the $\mathrm{K^{*0}/K}$ ratio in
panels (a) and (b) for different collision systems at
RHIC~\cite{Aggarwal:2010mt,Adams:2004ep,Adler:2002sw,Abelev:2008zk,Abelev:2008aa,Abelev:2012hy}
and at the
LHC~\cite{Adam:2017zbf,Abelev:2014uua,Abelev:2008yz,Adam:2016bpr,Acharya:2018orn}
as a function  of \avdndeta$^{1/3}$ and \snn, respectively. The
$\mathrm{K^{*0}/K}$ ratio in heavy-ion collisions is smaller than
those in \pp collisions, with the results from \pPb lying in
between. The $\mathrm{K^{*0}/K}$ ratio decreases when the system size
increases,  as reflected by the values of \avdndeta$^{1/3}$ (a proxy
for system size~\cite{Aamodt:2011mr}). To quantify the
suppression of $\mathrm{K^{*0}/K}$ ratio in central \PbPb collisions
with respect to \pp collisions, we calculate the double ratio $\left(
  \mathrm{K^{*0}/K} \right)_{\rm{PbPb}} / \left( \mathrm{K^{*0}/K}
\right)_{\rm{pp}}$. The $\mathrm{K^{*0}/K}$ double ratio in Pb--Pb
collisions at 5.02 TeV (2.76 TeV) is 0.483$\pm$0.082
(0.585$\pm$0.122), which deviates from unity by 6.2 (3.4) times its
standard deviation. The same ratio in Au--Au collisions at
\snn~$=$~200~GeV gives 0.571$\pm$0.147, which deviates from unity by
2.9 times its standard deviation.
Panels (b) of Fig.~\ref{fig:kstarphikratio} shows $\mathrm{K^{*0}/K}$ ratio 
as a function of \snn for \pp collisions, as well as for
central \pPb, \CuCu, \AuAu, and \PbPb collisions.
The
$\mathrm{K^{*0}/K}$ ratio is higher in \pp collisions than in central
\AuAu and \PbPb collisions at various center of mass energies. The
value of the $\mathrm{K^{*0}/K}$ ratio is larger in central \CuCu than
in central \AuAu collisions, as expected because of the smaller \CuCu system size.
  
Panels (c) and (d) of Fig.~\ref{fig:kstarphikratio} show the
$\mathrm{\phi/K}$ ratio  for different collision systems at RHIC~\cite{Aggarwal:2010mt,Adams:2004ep,Adler:2002sw,Abelev:2008zk,Abelev:2008aa,Abelev:2012hy} 
and
LHC~\cite{Adam:2017zbf,Abelev:2014uua,Abelev:2008yz,Adam:2016bpr,Acharya:2018orn}
as a function of \avdndeta$^{1/3}$ and \snn, respectively. In contrast
to the $\mathrm{K^{*0}/K}$ ratio, the $\mathrm{\phi/K}$ ratio is
approximately constant as a function of \avdndeta$^{1/3}$. 
The values of the $\mathrm{\phi/K}$~ratio in \AuAu and \CuCu
collisions are slightly larger than the corresponding results from
\PbPb collisions, but agree within uncertainties. The
$\mathrm{\phi/K}$ ratio is found to be independent of collision energy
and system from RHIC to LHC energies.

Figure~\ref{fig:kstarphikratio} (panels (a) and (c)) also shows the
$\mathrm{K^{*0}/K}$ and $\mathrm{\phi/K}$ ratios from EPOS3 model calculations with and without a
hadronic cascade  phase modeled by UrQMD~\cite{Knospe:2015nva}, and thermal model calculations with chemical freezeout temperature
$T_{\mathrm {ch}}$ $=$ 156 MeV~\cite{Stachel:2013zma}.
The thermal or statistical hadronization model assumes that the system
formed in heavy-ion collisions reaches thermal equilibrium through
multiple interactions and undergoes a rapid expansion followed by the
chemical freezeout. The freezeout surface is characterized by three
parameters: the chemical freezeout temperature $T_{\mathrm {ch}}$, the
chemical potential $\mu$ and the fireball volume V. The value of the
$\mathrm{K^{*0}/K}$ ratio in central \PbPb collisions is smaller than the thermal model expectation, however $\mathrm{\phi/K}$ ratio is in fair agreement with the model calculations. The EPOS3 event generator is based on 3$+$1D viscous hydrodynamical evolution where the initial stage is treated via multiple scattering approach based on Pomerons and strings and the reaction volume is divided into two parts, ``core'' and ``corona''.
It is the core part that provides the initial condition for QGP evolution, described by viscous hydrodynamics.
The corona part is composed of hadrons from the string decays. In EPOS3$+$UrQMD approach~\cite{Knospe:2015nva},
the hadrons separately produced from core and corona parts are fed into UrQMD~\cite{Bass:1998ca,Bleicher:1999xi}, which describes the hadronic interactions in a microscopic approach.
The chemical and kinetic freezeouts occur during this phase. The model predictions from EPOS3 and EPOS3$+$UrQMD are shown for \PbPb  collisions at \snn~$=$~2.76~TeV. As the ratios are shown as a function of \avdndeta,~no  significant qualitative differences are expected between the two energies. The observed trends of the $\mathrm{K^{*0}/K}$ and $\mathrm{\phi/K}$ ratios are reproduced by the EPOS3 generator with UrQMD. However, EPOS3 model without hadronic interactions is unable to reproduce the suppression of $\mathrm{K^{*0}/K}$ ratios towards the higher \avdndeta$^{1/3}$ values or central collisions.

\subsection{Nuclear modification factor}\label{sec:RAA}
The nuclear modification factor, $R_\mathrm{AA}$, of \kstar and \phim
mesons are studied as a function of centrality and center-of-mass
energy. The \RAA values of resonances are also compared to those of
$\pi$, K, and p to investigate the hadron species dependence of \RAA. 

\subsubsection{Centrality dependence of the nuclear modification factor}
The centrality dependence of \RAA helps in understanding the evolution of
parton energy loss in the medium as a function of the system size. 
\begin{figure}[tb]
    \begin{center}
      \includegraphics[width =  0.9\textwidth]{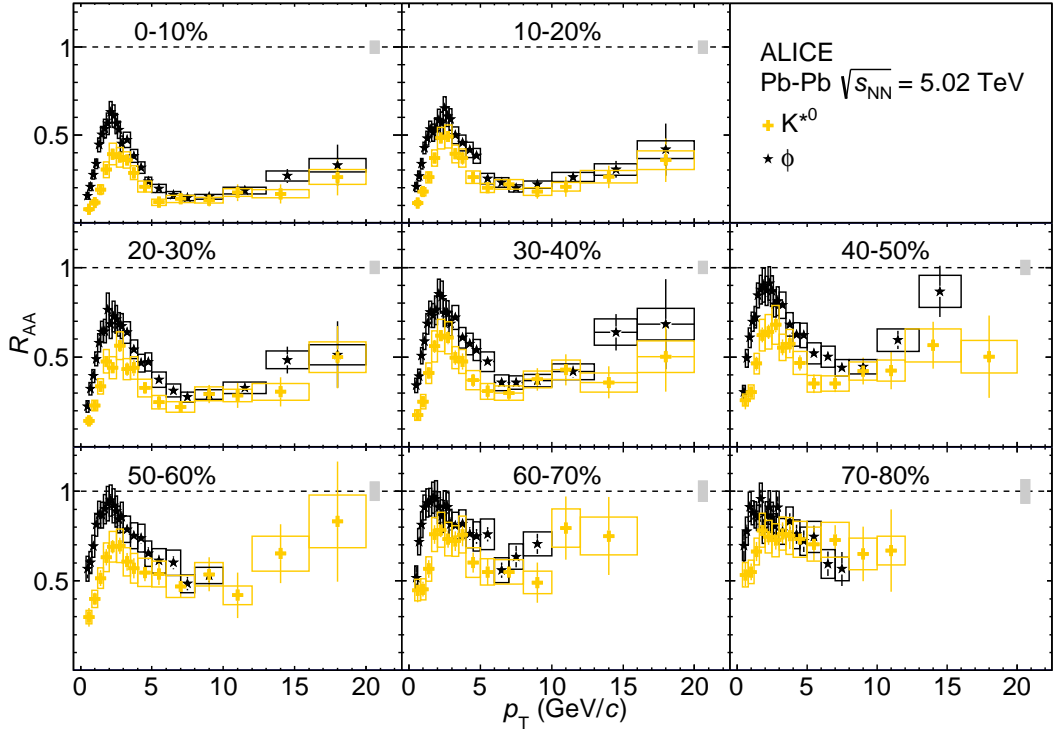}
    \end{center}
    \caption{The nuclear modification factor, \RAA, as a function of
      \pt for \kstar and \phim mesons at midrapidity ($|y| <$ 0.5)
      for different centrality classes in \PbPb  
      collisions at \snn $=$ 5.02 \TeV. The statistical and systematic
      uncertainties are shown as bars and boxes, respectively. The
      boxes around unity indicate the uncertainty on the normalization
      of \RAA.}
    \label{fig:RAACenDep}
  \end{figure}
Figure~\ref{fig:RAACenDep} shows the \RAA for \kstar and \phim mesons
as a function of \pt for different centrality classes at midrapidity
($|y| <$ 0.5) for \PbPb collisions at \snn $=$ 5.02 \TeV.  
The \RAA values are lower  for \kstar compared to \phim for \pt $<$ 5
GeV$/c$ for most of the collision centralities studied.  This can be
attributed to the dominance of rescattering effects at lower \pt. 
 At higher \pt ($>$ 6 GeV$/c$) the \RAA values for \kstar and \phim
 mesons are comparable within uncertainties.  The \RAA values below
 unity at high \pt  support the picture of a suppression of high \pt
 hadron production due to parton energy loss in the medium formed in
 heavy-ion collisions.  For all the collision centralities studied,
 the \RAA values are below unity and the values increase for \pt $>$ 6 GeV$/c$.
 The average \RAA values at high \pt ($>$ 6 GeV$/c$) are found to
 decrease when going from peripheral to central collisions  for both
 \kstar and \phim mesons. The dependence of \RAA on collision
 centrality at high \pt provides information on the path length
 dependence of parton energy loss in the medium formed in high energy heavy-ion
collisions~\cite{Adam:2015kca,Adam:2017zbf,Adare:2012wg,Christiansen:2013hya,Ortiz:2017cul,Acharya:2018eaq}.
This is reflected as a more pronounced suppression of \RAA in the most
central collisions, as expected from the longer path length traversed
by the hard partons as they  lose energy via multiple interactions.

\subsubsection{Center-of-mass energy dependence of the nuclear modification factor}
\begin{figure}[tb]
    \begin{center}
    \includegraphics[width = 0.9\textwidth]{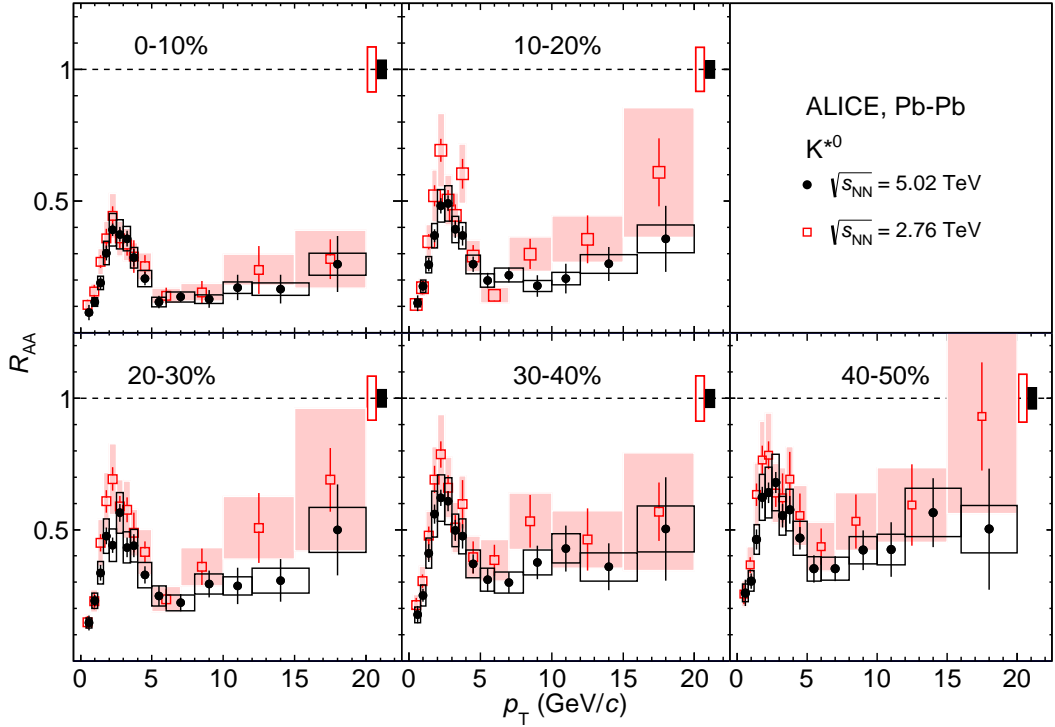}	
    \end{center}
    \caption{The nuclear modification factor, \RAA, as a function of
      \pt for \kstar for different centrality classes in \PbPb
      collisions at \snn $=$  5.02 \TeV compared with the results from
      \PbPb collisions at \snn $=$ 2.76 \TeV~\cite{Adam:2017zbf}. The statistical
      and systematic uncertainties are shown as bars and boxes,
      respectively. The boxes around unity indicate the respective
      uncertainty on the normalization of \RAA.}
    \label{fig:RAAKstarEnergyDep}
\end{figure}

\begin{figure}[tb]
    \begin{center}
      \includegraphics[width = 0.9\textwidth]{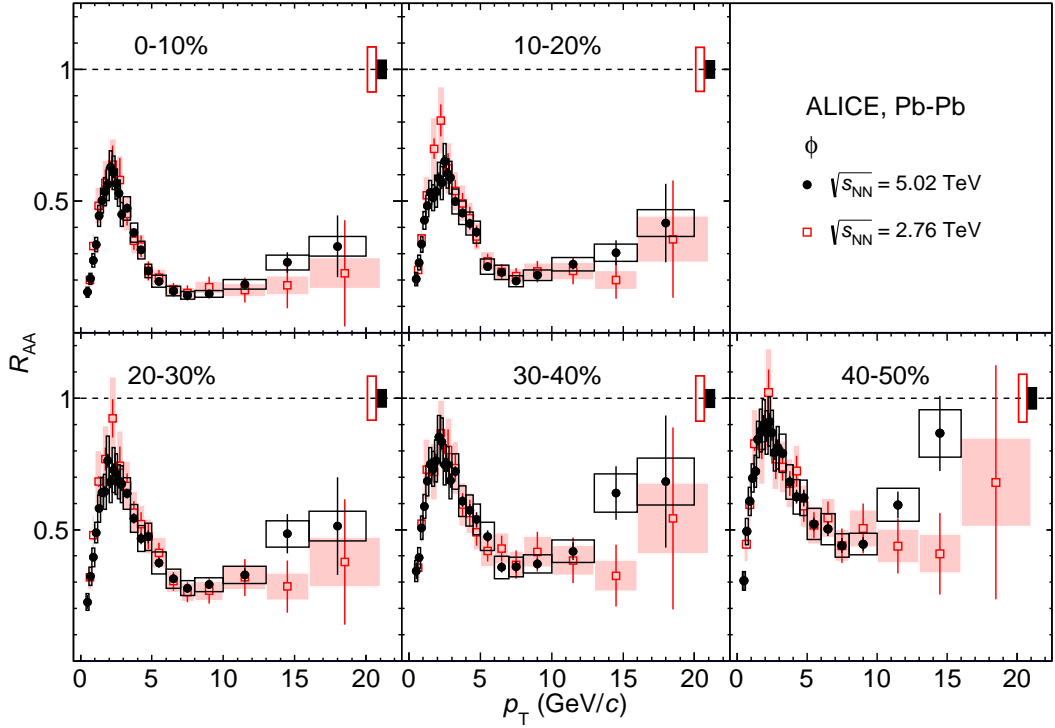}	
    \end{center}
    \caption{The nuclear modification factor, \RAA, as a function of
      \pt for \phim mesons in different centrality classes in \PbPb
      collisions at \snn $=$ 5.02 \TeV compared with the results in
      \PbPb collisions  at \snn $=$ 2.76 \TeV~\cite{Adam:2017zbf}. The
      statistical and systematic uncertainties are shown as bars and
      boxes, respectively. The boxes around unity indicate the
      uncertainty on the normalization of \RAA.}
    \label{fig:RAAphiEnergyDep}
  \end{figure}

Figure~\ref{fig:RAAKstarEnergyDep} (Figure~\ref{fig:RAAphiEnergyDep})
show the \pt-differential \RAA for the \kstar (\phim) meson for 5
different collision centrality classes (0--10\%, 10--20\%, 20--30\%,
30--40\%, and 40--50\%) in Pb--Pb collisions.  The \RAA values for \snn
$=$ 5.02 TeV are compared to the corresponding values at \snn $=$ 2.76
TeV.  No significant differences are observed between 5.02 TeV and 2.76
TeV for both the \kstar and \phim resonances. This is supported by
Figs.~\ref{fig:ptppen} and~\ref{fig:ptPbPben}, which show that the
ratios of \pt spectra of \kstar and \phim at 5.02 TeV to that at 2.76
TeV are similar for both pp and \PbPb collisions. The results from
Figs.~\ref{fig:RAACenDep},~\ref{fig:RAAKstarEnergyDep}
and~\ref{fig:RAAphiEnergyDep} indicate that within uncertainties, the
nuclear modification factor is independent of the resonance species
at high \pt and compatible with measurements at \snn $=$ 2.76 TeV for
different centrality classes. However, the \RAA measurements of other
mesonic and baryonic resonances like $\rho(770)^{0}$, $\Delta(1232)^{++}$, $\Sigma(1385)$, and $\Lambda(1520)$ which differ in lifetime, mass, quark content, and particle type are required to further support these results.

\subsubsection{Hadron species dependence of the nuclear modification factor}
\begin{figure}[tb]
    \begin{center}
    \includegraphics[width = 0.9\textwidth]{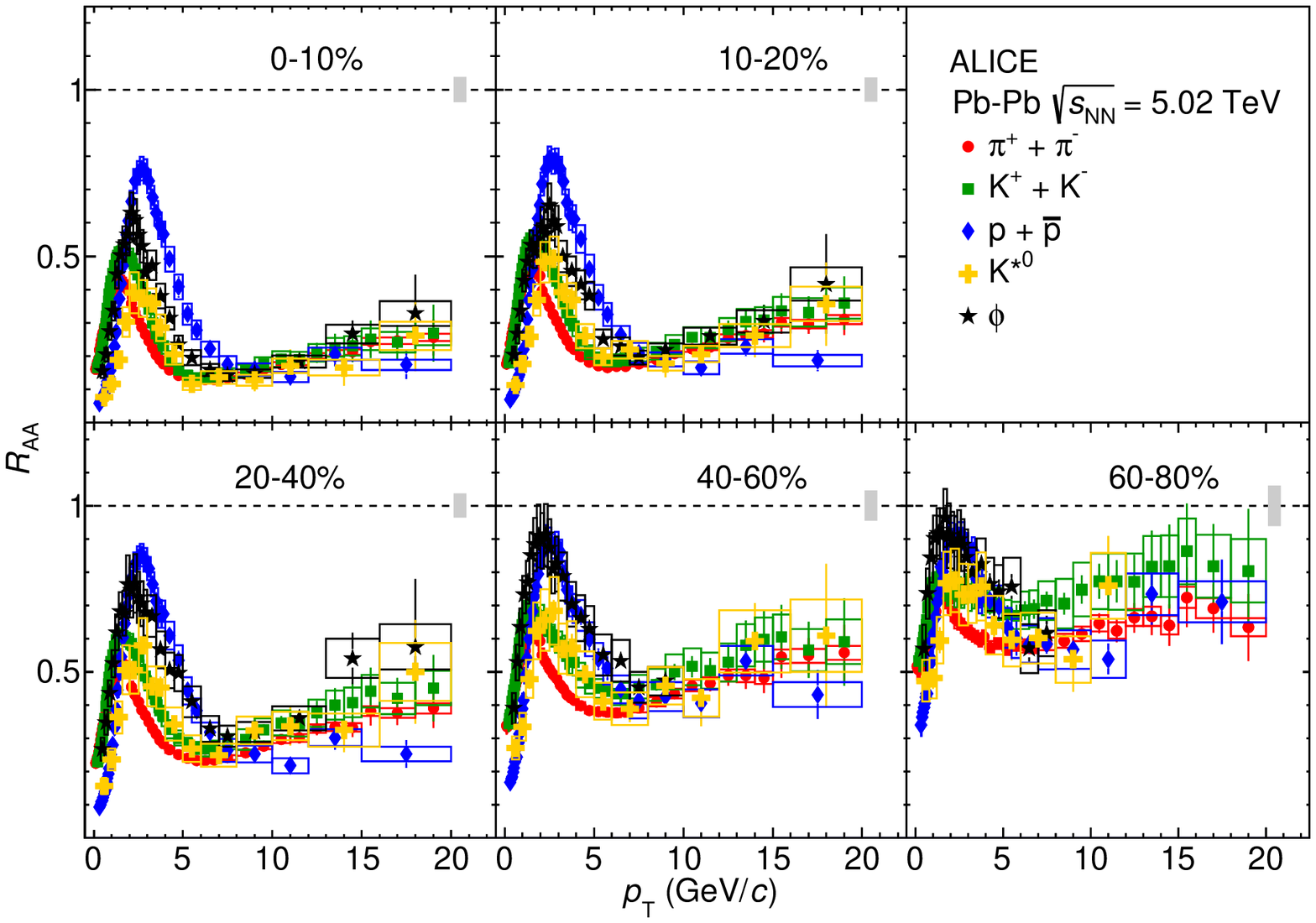}	
    \end{center}
    \caption{The nuclear modification factor, \RAA, as a function of
      \pt for \kstar and \phim mesons in \PbPb collisions for
      different centrality classes at \snn $=$ 5.02 \TeV. The results
      are compared with the \RAA values of $\pi$, K, and p measured by
      ALICE~\cite{Acharya:2019yoi}. The statistical and systematic
      uncertainties are shown as bars and boxes, respectively. The
      boxes around unity indicate the uncertainty on the normalization
      of \RAA.} \label{fig:RAAEnergyDep}
  \end{figure}
Figure~\ref{fig:RAAEnergyDep} shows the hadron species dependence of
\RAA for various collision centrality classes in \PbPb collisions at
\snn $=$ 5.02 TeV. The hadron species considered here are charged
pions, kaons, (anti)protons, \kstar and \phim mesons. They vary in
mass from about 140 MeV to 1019 MeV, both baryons and mesons are
considered, and their valence quark contents are also different. At low
\pt ($<$ 2 GeV$/c$), the \kstar \RAA values are the smallest for
central collisions; this is attributed to the rescattering effect as
discussed earlier in the article. For the \pt range 2-8 GeV$/c$, there
appears to be a hadron mass dependence for mesons. It is also observed
that \RAA of proton is higher than all mesons including \phim. This
indicate a baryon-meson ordering. So, even in the presence of strong
radial  flow in this \pt region, there are other effects which can
affect \RAA. For \pt larger than 8 GeV$/c$, all the particle species
show similar \RAA values within  uncertainties for all the collision
centralities studied. This suggests that, despite the varying degree
of energy loss at different collision centralities, the relative
particle composition at high \pt remains the same as in vacuum.    

\section{Conclusions}\label{sec:conc}
The transverse momentum distributions of \kstar and \phim mesons 
have been measured in inelastic \pp and \PbPb collisions at \snn $=$
5.02 TeV using the ALICE detector. The measurements are carried out at
midrapidity ($|y| <$ 0.5) up to \pt $=$ 20 GeV$/c$ and for \PbPb
collisions in various collision centrality classes.  

The \pt distributions for \kstar and \phim mesons in \pp collisions
are well described by a  L\'{e}vy-Tsallis function and are compared to
results from the PYTHIA6, PYTHIA8, EPOS-LHC, and HERWIG event generators. 
None of the models are able to describe the transverse momentum distributions of \kstar and \phim in the measured \pt range. 
The \meanpt values are found to increase with
\avdndeta$_{|\eta|\textless0.5}$, with the mass of the hadron and with
\snn. These measurements are consistent with the observation that
radial flow effects are larger for more central collisions and
increased high-\pt production at higher collision energies.

The \pt-integrated particle ratios as a function of \snn and \avdndeta$^{1/3}$ in Pb--Pb and inelastic pp collisions have been compared. The $\mathrm{K^{*0}/K}$ and $\mathrm{\phi/K}$ ratios as a function of \snn and \avdndeta$^{1/3}$ taken together indicate the dominance of the rescattering effect in the hadronic phase in \PbPb collisions. EPOS3 and thermal models that do not include hadronic interactions are unable to describe the suppression of $\mathrm{K^{*0}/K}$ ratio. EPOS3$+$UrQMD, where the hadronic phase is described by the UrQMD model, is able to reproduce the decreasing trend of $\mathrm{K^{*0}/K}$ ratio as a function of multiplicity in the \PbPb collisions. In contrast, the $\mathrm{\phi/K}$ ratios in \PbPb collisions are quite comparable to those from \pp collisions, and agree well with all model calculations.  The dissimilarity in the behavior of $\mathrm{K^{*0}/K}$ and $\mathrm{\phi/K}$ ratios is dominantly attributed to the lifetime of \kstar, which is a factor of 10 smaller
than the lifetime of the \phim meson. Hence, \kstar decay daughters are subjected to a greater rescattering in the hadronic medium.
The comparison of transverse momentum distributions of  \kstar and \phim in central \PbPb collisions with blast-wave predictions,
which does not include rescattering effects, show a suppression of \kstar yield for \pt$<$ 3 GeV$/c$.

At low \pt ($<$ 5 GeV$/c$), the nuclear modification factor values for \kstar are lower with respect to those obtained for \phim mesons, chiefly because of rescattering effects.
The \RAA values at high \pt ($>$ 6--8 GeV$/c$) for \kstar
and \phim mesons are comparable within uncertainties. The average \RAA
values at high \pt are found to decrease when going from peripheral to central
collisions. The \RAA values at high \pt for \kstar and \phim mesons at
\snn~$=$~5.02~\TeV  are comparable to the corresponding measurements
at \snn~$=$~2.76~TeV for most of the collision centralities studied.
At the same time, the transverse momentum spectra at high \pt in both
\pp and \PbPb collisions are found to be higher by a  factor of about
1.8 at \snn~$=$~5.02~TeV compared to \snn~$=$~2.76~TeV.  
 Further, we find that the \RAA values at high \pt for the hadrons
$\pi$, K, \kstar, p, and \phim are similar within  uncertainties for
all the collision centrality classes studied. This suggests that the
energy loss in the medium which leads to the suppression does not
modify the particle composition in the light quark sector. 


\newenvironment{acknowledgement}{\relax}{\relax}
\begin{acknowledgement}
\section*{Acknowledgements}

The ALICE Collaboration would like to thank all its engineers and technicians for their invaluable contributions to the construction of the experiment and the CERN accelerator teams for the outstanding performance of the LHC complex.
The ALICE Collaboration gratefully acknowledges the resources and support provided by all Grid centres and the Worldwide LHC Computing Grid (WLCG) collaboration.
The ALICE Collaboration acknowledges the following funding agencies for their support in building and running the ALICE detector:
A. I. Alikhanyan National Science Laboratory (Yerevan Physics Institute) Foundation (ANSL), State Committee of Science and World Federation of Scientists (WFS), Armenia;
Austrian Academy of Sciences, Austrian Science Fund (FWF): [M 2467-N36] and Nationalstiftung f\"{u}r Forschung, Technologie und Entwicklung, Austria;
Ministry of Communications and High Technologies, National Nuclear Research Center, Azerbaijan;
Conselho Nacional de Desenvolvimento Cient\'{\i}fico e Tecnol\'{o}gico (CNPq), Financiadora de Estudos e Projetos (Finep), Funda\c{c}\~{a}o de Amparo \`{a} Pesquisa do Estado de S\~{a}o Paulo (FAPESP) and Universidade Federal do Rio Grande do Sul (UFRGS), Brazil;
Ministry of Education of China (MOEC) , Ministry of Science \& Technology of China (MSTC) and National Natural Science Foundation of China (NSFC), China;
Ministry of Science and Education and Croatian Science Foundation, Croatia;
Centro de Aplicaciones Tecnol\'{o}gicas y Desarrollo Nuclear (CEADEN), Cubaenerg\'{\i}a, Cuba;
Ministry of Education, Youth and Sports of the Czech Republic, Czech Republic;
The Danish Council for Independent Research | Natural Sciences, the VILLUM FONDEN and Danish National Research Foundation (DNRF), Denmark;
Helsinki Institute of Physics (HIP), Finland;
Commissariat \`{a} l'Energie Atomique (CEA) and Institut National de Physique Nucl\'{e}aire et de Physique des Particules (IN2P3) and Centre National de la Recherche Scientifique (CNRS), France;
Bundesministerium f\"{u}r Bildung und Forschung (BMBF) and GSI Helmholtzzentrum f\"{u}r Schwerionenforschung GmbH, Germany;
General Secretariat for Research and Technology, Ministry of Education, Research and Religions, Greece;
National Research, Development and Innovation Office, Hungary;
Department of Atomic Energy Government of India (DAE), Department of Science and Technology, Government of India (DST), University Grants Commission, Government of India (UGC) and Council of Scientific and Industrial Research (CSIR), India;
Indonesian Institute of Science, Indonesia;
Istituto Nazionale di Fisica Nucleare (INFN), Italy;
Institute for Innovative Science and Technology , Nagasaki Institute of Applied Science (IIST), Japanese Ministry of Education, Culture, Sports, Science and Technology (MEXT) and Japan Society for the Promotion of Science (JSPS) KAKENHI, Japan;
Consejo Nacional de Ciencia (CONACYT) y Tecnolog\'{i}a, through Fondo de Cooperaci\'{o}n Internacional en Ciencia y Tecnolog\'{i}a (FONCICYT) and Direcci\'{o}n General de Asuntos del Personal Academico (DGAPA), Mexico;
Nederlandse Organisatie voor Wetenschappelijk Onderzoek (NWO), Netherlands;
The Research Council of Norway, Norway;
Commission on Science and Technology for Sustainable Development in the South (COMSATS), Pakistan;
Pontificia Universidad Cat\'{o}lica del Per\'{u}, Peru;
Ministry of Education and Science, National Science Centre and WUT ID-UB, Poland;
Korea Institute of Science and Technology Information and National Research Foundation of Korea (NRF), Republic of Korea;
Ministry of Education and Scientific Research, Institute of Atomic Physics and Ministry of Research and Innovation and Institute of Atomic Physics, Romania;
Joint Institute for Nuclear Research (JINR), Ministry of Education and Science of the Russian Federation, National Research Centre Kurchatov Institute, Russian Science Foundation and Russian Foundation for Basic Research, Russia;
Ministry of Education, Science, Research and Sport of the Slovak Republic, Slovakia;
National Research Foundation of South Africa, South Africa;
Swedish Research Council (VR) and Knut \& Alice Wallenberg Foundation (KAW), Sweden;
European Organization for Nuclear Research, Switzerland;
Suranaree University of Technology (SUT), National Science and Technology Development Agency (NSDTA) and Office of the Higher Education Commission under NRU project of Thailand, Thailand;
Turkish Energy, Nuclear and Mineral Research Agency (TENMAK), Turkey;
National Academy of  Sciences of Ukraine, Ukraine;
Science and Technology Facilities Council (STFC), United Kingdom;
National Science Foundation of the United States of America (NSF) and United States Department of Energy, Office of Nuclear Physics (DOE NP), United States of America.
\end{acknowledgement}

\bibliographystyle{utphys}   
\bibliography{KStarPhiPbPb5p02TeV_longpaper}

\newpage
\appendix

%
%

\section{The ALICE Collaboration}
\label{app:collab}

\small
\begin{flushleft}
S.~Acharya$^{\rm 143}$, 
D.~Adamov\'{a}$^{\rm 98}$, 
A.~Adler$^{\rm 76}$, 
J.~Adolfsson$^{\rm 83}$, 
G.~Aglieri Rinella$^{\rm 35}$, 
M.~Agnello$^{\rm 31}$, 
N.~Agrawal$^{\rm 55}$, 
Z.~Ahammed$^{\rm 143}$, 
S.~Ahmad$^{\rm 16}$, 
S.U.~Ahn$^{\rm 78}$
% ...

\section*{Affiliation notes}

$^{\rm I}$ Deceased\\
$^{\rm II}$ Also at: Italian National Agency for New Technologies, Energy and Sustainable Economic Development (ENEA), Bologna, Italy\\
$^{\rm III}$ Also at: Dipartimento DET del Politecnico di Torino, Turin, Italy\\
$^{\rm IV}$ Also at: M.V. Lomonosov Moscow State University, D.V. Skobeltsyn Institute of Nuclear, Physics, Moscow, Russia\\
$^{\rm V}$ Also at: Institute of Theoretical Physics, University of Wroclaw, Poland\\

\section*{Collaboration Institutes}

$^{1}$ A.I. Alikhanyan National Science Laboratory (Yerevan Physics Institute) Foundation, Yerevan, Armenia\\
$^{2}$ AGH University of Science and Technology, Cracow, Poland\\
$^{3}$ Bogolyubov Institute for Theoretical Physics, National Academy of Sciences of Ukraine, Kiev, Ukraine\\
$^{4}$ Bose Institute, Department of Physics  and Centre for Astroparticle Physics and Space Science (CAPSS), Kolkata, India\\
$^{5}$ Budker Institute for Nuclear Physics, Novosibirsk, Russia\\
$^{6}$ California Polytechnic State University, San Luis Obispo, California, United States\\
$^{7}$ Central China Normal University, Wuhan, China\\
$^{8}$ Centro de Aplicaciones Tecnol\'{o}gicas y Desarrollo Nuclear (CEADEN), Havana, Cuba\\
$^{9}$ Centro de Investigaci\'{o}n y de Estudios Avanzados (CINVESTAV), Mexico City and M\'{e}rida, Mexico\\
$^{10}$ Chicago State University, Chicago, Illinois, United States\\
$^{11}$ China Institute of Atomic Energy, Beijing, China\\
$^{12}$ Chungbuk National University, Cheongju, Republic of Korea\\
$^{13}$ Comenius University Bratislava, Faculty of Mathematics, Physics and Informatics, Bratislava, Slovakia\\
$^{14}$ COMSATS University Islamabad, Islamabad, Pakistan\\
$^{15}$ Creighton University, Omaha, Nebraska, United States\\
$^{16}$ Department of Physics, Aligarh Muslim University, Aligarh, India\\
$^{17}$ Department of Physics, Pusan National University, Pusan, Republic of Korea\\
$^{18}$ Department of Physics, Sejong University, Seoul, Republic of Korea\\
$^{19}$ Department of Physics, University of California, Berkeley, California, United States\\
$^{20}$ Department of Physics, University of Oslo, Oslo, Norway\\
$^{21}$ Department of Physics and Technology, University of Bergen, Bergen, Norway\\
$^{22}$ Dipartimento di Fisica dell'Universit\`{a} 'La Sapienza' and Sezione INFN, Rome, Italy\\
$^{23}$ Dipartimento di Fisica dell'Universit\`{a} and Sezione INFN, Cagliari, Italy\\
$^{24}$ Dipartimento di Fisica dell'Universit\`{a} and Sezione INFN, Trieste, Italy\\
$^{25}$ Dipartimento di Fisica dell'Universit\`{a} and Sezione INFN, Turin, Italy\\
$^{26}$ Dipartimento di Fisica e Astronomia dell'Universit\`{a} and Sezione INFN, Bologna, Italy\\
$^{27}$ Dipartimento di Fisica e Astronomia dell'Universit\`{a} and Sezione INFN, Catania, Italy\\
$^{28}$ Dipartimento di Fisica e Astronomia dell'Universit\`{a} and Sezione INFN, Padova, Italy\\
$^{29}$ Dipartimento di Fisica e Nucleare e Teorica, Universit\`{a} di Pavia, Pavia, Italy\\
$^{30}$ Dipartimento di Fisica `E.R.~Caianiello' dell'Universit\`{a} and Gruppo Collegato INFN, Salerno, Italy\\
$^{31}$ Dipartimento DISAT del Politecnico and Sezione INFN, Turin, Italy\\
$^{32}$ Dipartimento di Scienze e Innovazione Tecnologica dell'Universit\`{a} del Piemonte Orientale and INFN Sezione di Torino, Alessandria, Italy\\
$^{33}$ Dipartimento di Scienze MIFT, Universit\`{a} di Messina, Messina, Italy\\
$^{34}$ Dipartimento Interateneo di Fisica `M.~Merlin' and Sezione INFN, Bari, Italy\\
$^{35}$ European Organization for Nuclear Research (CERN), Geneva, Switzerland\\
$^{36}$ Faculty of Electrical Engineering, Mechanical Engineering and Naval Architecture, University of Split, Split, Croatia\\
$^{37}$ Faculty of Engineering and Science, Western Norway University of Applied Sciences, Bergen, Norway\\
$^{38}$ Faculty of Nuclear Sciences and Physical Engineering, Czech Technical University in Prague, Prague, Czech Republic\\
$^{39}$ Faculty of Science, P.J.~\v{S}af\'{a}rik University, Ko\v{s}ice, Slovakia\\
$^{40}$ Frankfurt Institute for Advanced Studies, Johann Wolfgang Goethe-Universit\"{a}t Frankfurt, Frankfurt, Germany\\
$^{41}$ Fudan University, Shanghai, China\\
$^{42}$ Gangneung-Wonju National University, Gangneung, Republic of Korea\\
$^{43}$ Gauhati University, Department of Physics, Guwahati, India\\
$^{44}$ Helmholtz-Institut f\"{u}r Strahlen- und Kernphysik, Rheinische Friedrich-Wilhelms-Universit\"{a}t Bonn, Bonn, Germany\\
$^{45}$ Helsinki Institute of Physics (HIP), Helsinki, Finland\\
$^{46}$ High Energy Physics Group,  Universidad Aut\'{o}noma de Puebla, Puebla, Mexico\\
$^{47}$ Hiroshima University, Hiroshima, Japan\\
$^{48}$ Hochschule Worms, Zentrum  f\"{u}r Technologietransfer und Telekommunikation (ZTT), Worms, Germany\\
$^{49}$ Horia Hulubei National Institute of Physics and Nuclear Engineering, Bucharest, Romania\\
$^{50}$ Indian Institute of Technology Bombay (IIT), Mumbai, India\\
$^{51}$ Indian Institute of Technology Indore, Indore, India\\
$^{52}$ Indonesian Institute of Sciences, Jakarta, Indonesia\\
$^{53}$ INFN, Laboratori Nazionali di Frascati, Frascati, Italy\\
$^{54}$ INFN, Sezione di Bari, Bari, Italy\\
$^{55}$ INFN, Sezione di Bologna, Bologna, Italy\\
$^{56}$ INFN, Sezione di Cagliari, Cagliari, Italy\\
$^{57}$ INFN, Sezione di Catania, Catania, Italy\\
$^{58}$ INFN, Sezione di Padova, Padova, Italy\\
$^{59}$ INFN, Sezione di Pavia, Pavia, Italy\\
$^{60}$ INFN, Sezione di Roma, Rome, Italy\\
$^{61}$ INFN, Sezione di Torino, Turin, Italy\\
$^{62}$ INFN, Sezione di Trieste, Trieste, Italy\\
$^{63}$ Inha University, Incheon, Republic of Korea\\
$^{64}$ Institute for Gravitational and Subatomic Physics (GRASP), Utrecht University/Nikhef, Utrecht, Netherlands\\
$^{65}$ Institute for Nuclear Research, Academy of Sciences, Moscow, Russia\\
$^{66}$ Institute of Experimental Physics, Slovak Academy of Sciences, Ko\v{s}ice, Slovakia\\
$^{67}$ Institute of Physics, Homi Bhabha National Institute, Bhubaneswar, India\\
$^{68}$ Institute of Physics of the Czech Academy of Sciences, Prague, Czech Republic\\
$^{69}$ Institute of Space Science (ISS), Bucharest, Romania\\
$^{70}$ Institut f\"{u}r Kernphysik, Johann Wolfgang Goethe-Universit\"{a}t Frankfurt, Frankfurt, Germany\\
$^{71}$ Instituto de Ciencias Nucleares, Universidad Nacional Aut\'{o}noma de M\'{e}xico, Mexico City, Mexico\\
$^{72}$ Instituto de F\'{i}sica, Universidade Federal do Rio Grande do Sul (UFRGS), Porto Alegre, Brazil\\
$^{73}$ Instituto de F\'{\i}sica, Universidad Nacional Aut\'{o}noma de M\'{e}xico, Mexico City, Mexico\\
$^{74}$ iThemba LABS, National Research Foundation, Somerset West, South Africa\\
$^{75}$ Jeonbuk National University, Jeonju, Republic of Korea\\
$^{76}$ Johann-Wolfgang-Goethe Universit\"{a}t Frankfurt Institut f\"{u}r Informatik, Fachbereich Informatik und Mathematik, Frankfurt, Germany\\
$^{77}$ Joint Institute for Nuclear Research (JINR), Dubna, Russia\\
$^{78}$ Korea Institute of Science and Technology Information, Daejeon, Republic of Korea\\
$^{79}$ KTO Karatay University, Konya, Turkey\\
$^{80}$ Laboratoire de Physique des 2 Infinis, Ir\`{e}ne Joliot-Curie, Orsay, France\\
$^{81}$ Laboratoire de Physique Subatomique et de Cosmologie, Universit\'{e} Grenoble-Alpes, CNRS-IN2P3, Grenoble, France\\
$^{82}$ Lawrence Berkeley National Laboratory, Berkeley, California, United States\\
$^{83}$ Lund University Department of Physics, Division of Particle Physics, Lund, Sweden\\
$^{84}$ Moscow Institute for Physics and Technology, Moscow, Russia\\
$^{85}$ Nagasaki Institute of Applied Science, Nagasaki, Japan\\
$^{86}$ Nara Women{'}s University (NWU), Nara, Japan\\
$^{87}$ National and Kapodistrian University of Athens, School of Science, Department of Physics , Athens, Greece\\
$^{88}$ National Centre for Nuclear Research, Warsaw, Poland\\
$^{89}$ National Institute of Science Education and Research, Homi Bhabha National Institute, Jatni, India\\
$^{90}$ National Nuclear Research Center, Baku, Azerbaijan\\
$^{91}$ National Research Centre Kurchatov Institute, Moscow, Russia\\
$^{92}$ Niels Bohr Institute, University of Copenhagen, Copenhagen, Denmark\\
$^{93}$ Nikhef, National institute for subatomic physics, Amsterdam, Netherlands\\
$^{94}$ NRC Kurchatov Institute IHEP, Protvino, Russia\\
$^{95}$ NRC \guillemotleft Kurchatov\guillemotright  Institute - ITEP, Moscow, Russia\\
$^{96}$ NRNU Moscow Engineering Physics Institute, Moscow, Russia\\
$^{97}$ Nuclear Physics Group, STFC Daresbury Laboratory, Daresbury, United Kingdom\\
$^{98}$ Nuclear Physics Institute of the Czech Academy of Sciences, \v{R}e\v{z} u Prahy, Czech Republic\\
$^{99}$ Oak Ridge National Laboratory, Oak Ridge, Tennessee, United States\\
$^{100}$ Ohio State University, Columbus, Ohio, United States\\
$^{101}$ Petersburg Nuclear Physics Institute, Gatchina, Russia\\
$^{102}$ Physics department, Faculty of science, University of Zagreb, Zagreb, Croatia\\
$^{103}$ Physics Department, Panjab University, Chandigarh, India\\
$^{104}$ Physics Department, University of Jammu, Jammu, India\\
$^{105}$ Physics Department, University of Rajasthan, Jaipur, India\\
$^{106}$ Physikalisches Institut, Eberhard-Karls-Universit\"{a}t T\"{u}bingen, T\"{u}bingen, Germany\\
$^{107}$ Physikalisches Institut, Ruprecht-Karls-Universit\"{a}t Heidelberg, Heidelberg, Germany\\
$^{108}$ Physik Department, Technische Universit\"{a}t M\"{u}nchen, Munich, Germany\\
$^{109}$ Politecnico di Bari and Sezione INFN, Bari, Italy\\
$^{110}$ Research Division and ExtreMe Matter Institute EMMI, GSI Helmholtzzentrum f\"ur Schwerionenforschung GmbH, Darmstadt, Germany\\
$^{111}$ Russian Federal Nuclear Center (VNIIEF), Sarov, Russia\\
$^{112}$ Saha Institute of Nuclear Physics, Homi Bhabha National Institute, Kolkata, India\\
$^{113}$ School of Physics and Astronomy, University of Birmingham, Birmingham, United Kingdom\\
$^{114}$ Secci\'{o}n F\'{\i}sica, Departamento de Ciencias, Pontificia Universidad Cat\'{o}lica del Per\'{u}, Lima, Peru\\
$^{115}$ St. Petersburg State University, St. Petersburg, Russia\\
$^{116}$ Stefan Meyer Institut f\"{u}r Subatomare Physik (SMI), Vienna, Austria\\
$^{117}$ SUBATECH, IMT Atlantique, Universit\'{e} de Nantes, CNRS-IN2P3, Nantes, France\\
$^{118}$ Suranaree University of Technology, Nakhon Ratchasima, Thailand\\
$^{119}$ Technical University of Ko\v{s}ice, Ko\v{s}ice, Slovakia\\
$^{120}$ The Henryk Niewodniczanski Institute of Nuclear Physics, Polish Academy of Sciences, Cracow, Poland\\
$^{121}$ The University of Texas at Austin, Austin, Texas, United States\\
$^{122}$ Universidad Aut\'{o}noma de Sinaloa, Culiac\'{a}n, Mexico\\
$^{123}$ Universidade de S\~{a}o Paulo (USP), S\~{a}o Paulo, Brazil\\
$^{124}$ Universidade Estadual de Campinas (UNICAMP), Campinas, Brazil\\
$^{125}$ Universidade Federal do ABC, Santo Andre, Brazil\\
$^{126}$ University of Cape Town, Cape Town, South Africa\\
$^{127}$ University of Houston, Houston, Texas, United States\\
$^{128}$ University of Jyv\"{a}skyl\"{a}, Jyv\"{a}skyl\"{a}, Finland\\
$^{129}$ University of Kansas, Lawrence, Kansas, United States\\
$^{130}$ University of Liverpool, Liverpool, United Kingdom\\
$^{131}$ University of Science and Technology of China, Hefei, China\\
$^{132}$ University of South-Eastern Norway, Tonsberg, Norway\\
$^{133}$ University of Tennessee, Knoxville, Tennessee, United States\\
$^{134}$ University of the Witwatersrand, Johannesburg, South Africa\\
$^{135}$ University of Tokyo, Tokyo, Japan\\
$^{136}$ University of Tsukuba, Tsukuba, Japan\\
$^{137}$ Universit\'{e} Clermont Auvergne, CNRS/IN2P3, LPC, Clermont-Ferrand, France\\
$^{138}$ Universit\'{e} de Lyon, CNRS/IN2P3, Institut de Physique des 2 Infinis de Lyon , Lyon, France\\
$^{139}$ Universit\'{e} de Strasbourg, CNRS, IPHC UMR 7178, F-67000 Strasbourg, France, Strasbourg, France\\
$^{140}$ Universit\'{e} Paris-Saclay Centre d'Etudes de Saclay (CEA), IRFU, D\'{e}partment de Physique Nucl\'{e}aire (DPhN), Saclay, France\\
$^{141}$ Universit\`{a} degli Studi di Foggia, Foggia, Italy\\
$^{142}$ Universit\`{a} di Brescia, Brescia, Italy\\
$^{143}$ Variable Energy Cyclotron Centre, Homi Bhabha National Institute, Kolkata, India\\
$^{144}$ Warsaw University of Technology, Warsaw, Poland\\
$^{145}$ Wayne State University, Detroit, Michigan, United States\\
$^{146}$ Westf\"{a}lische Wilhelms-Universit\"{a}t M\"{u}nster, Institut f\"{u}r Kernphysik, M\"{u}nster, Germany\\
$^{147}$ Wigner Research Centre for Physics, Budapest, Hungary\\
$^{148}$ Yale University, New Haven, Connecticut, United States\\
$^{149}$ Yonsei University, Seoul, Republic of Korea\\

\bigskip 

\end{flushleft} 
  
\end{document}